\documentclass{article}
\usepackage{arxiv}
\usepackage[utf8]{inputenc}
\usepackage[T1]{fontenc}

\usepackage[square,numbers,sort&compress]{natbib}
\usepackage{graphicx}
\usepackage{amsmath}
\usepackage{amssymb}
\usepackage{amsfonts}
\usepackage{bm}
\usepackage{url}
\usepackage{booktabs}
\usepackage{nicefrac}
\usepackage{xcolor}
\usepackage{multirow}
\usepackage{comment}
\usepackage{colortbl}
\usepackage{wrapfig}
\usepackage{tabularx}
\usepackage{textcomp}
\usepackage{stfloats}
\usepackage{verbatim}
\usepackage{adjustbox}
\usepackage{pifont}
\usepackage{tikz}
\usepackage{subcaption}
\usepackage{caption}
\usepackage{float}
\usepackage{array}
\usepackage{makecell}
\usepackage{threeparttable}
\usepackage{enumitem}
\usepackage[normalem]{ulem}
\usepackage{microtype}
\usepackage{listings}
\usepackage[most]{tcolorbox}
\usepackage{hyperref}
\usepackage{cleveref}
\usepackage{xspace}

\makeatletter
\DeclareRobustCommand\onedot{\futurelet\@let@token\@onedot}
\def\@onedot{\ifx\@let@token.\else.\null\fi\xspace}
\makeatother

\definecolor{lightblue}{rgb}{0.66, 0.85, 0.95}
\definecolor{c2}{HTML}{FBD9BD}
\definecolor{c3}{HTML}{fe793d}
\definecolor{c4}{HTML}{eedeb0}
\definecolor{rouse}{rgb}{0.981,0.961,0.941}
\definecolor{adptorange}{RGB}{248, 205, 172}
\definecolor{cmpblue}{RGB}{189, 215, 238}
\definecolor{our_red}{RGB}{232,157,160}
\definecolor{our_blue}{RGB}{136,206,230}
\definecolor{our_orange}{RGB}{246,200,168}
\definecolor{our_green}{RGB}{178,211,164}
\definecolor{attn_code0}{RGB}{247,215,200}
\definecolor{attn_code1}{RGB}{238,169,139}
\definecolor{mlp_code0}{RGB}{204,201,221}
\definecolor{mlp_code1}{RGB}{102,95,153}
\definecolor{token_blue}{RGB}{84, 120, 140}
\definecolor{linkblue}{HTML}{003E7D}

\hypersetup{
    colorlinks=true,
    linkcolor=red,
    citecolor=blue,
    urlcolor=blue,
}

\tcbset{colback=blue!5!white, colframe=blue!20!white, boxrule=0pt, arc=2mm, left=1mm, right=1mm, top=1mm, bottom=1mm}

\newlength\savewidth

\newcolumntype{x}[1]{>{\centering\arraybackslash}p{#1pt}}
\newcolumntype{y}[1]{>{\raggedright\arraybackslash}p{#1pt}}
\newcolumntype{z}[1]{>{\raggedleft\arraybackslash}p{#1pt}}

\definecolor{codeblue}{rgb}{0.25, 0.5, 0.5}
\definecolor{codekw}{rgb}{0.35, 0.35, 0.75}
\lstdefinestyle{Pytorch}{
    language = Python,
    backgroundcolor = \color{white},
    basicstyle = \fontsize{9pt}{8pt}\selectfont\ttfamily\bfseries,
    columns = fullflexible,
    aboveskip=1pt,
    belowskip=1pt,
    breaklines = true,
    captionpos = b,
    commentstyle = \color{codeblue},
    keywordstyle = \color{codekw},
}

\definecolor{gain_green}{HTML}{009000}
\definecolor{loss_red}{HTML}{ea4335}

\setlist[itemize,1]{leftmargin=18pt}
\setlist[enumerate,1]{leftmargin=18pt}
\title{
\raisebox{-0.1\height}{\includegraphics[width=0.04\textwidth]{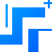}} %
Boosting Omni-Modal Language Models:\\ Staged Post-Training with Visually Debiased Evaluation
}

\author{\vspace{1em} StepFun-Audio Team\\
\vspace{1em}
\href{https://cheliu-computation.github.io/omni/}{Project Page}
\quad|\quad
\href{https://huggingface.co/datasets/che111/OmniClean}{OmniClean Dataset}
}

\renewcommand{\headeright}{StepFun-Audio Team}
\renewcommand{\undertitle}{}
\renewcommand{\shorttitle}{}

\begin{document}
\large
\maketitle

\begin{abstract}
Omni-modal language models are designed to jointly understand audio, visual inputs, and language, yet their benchmark gains do not necessarily reflect genuine omni-modal understanding: when visual evidence alone is sufficient, improvements can be driven by visual shortcuts rather than better omni-modal integration. We ask whether existing omni-modal benchmarks can separate such shortcuts from audio-visual-language evidence integration, and how post-training behaves under a visually debiased evaluation setting. To this end, we audit nine omni benchmarks with visual-only probing, remove visually solvable queries, and retain full subsets only when filtering is undefined or would destabilize score comparisons. This protocol audits 16,968 queries and yields \textbf{OmniClean}, a visually debiased evaluation view with 8,551 retained queries. On this testbed, we study \textbf{OmniBoost}, a three-stage post-training recipe based on Qwen2.5-Omni-3B: mixed bi-modal SFT, mixed-modality RLVR, and SFT on self-distilled data. The staged results show that balanced bi-modal SFT alone yields limited and uneven gains, whereas RLVR provides the first broad improvement and self-distillation further reshapes the benchmark profile. The competitive gains come from the staged post-training recipe and the synthetic-query construction: after SFT on self-distilled data, the 3B model becomes comparable to larger open-source references and slightly exceeds Qwen3-Omni-30B-A3B-Instruct under both OmniClean aggregate summaries, without distilling answers from a stronger omni-modal teacher. These findings suggest that omni-modal progress is more meaningfully assessed when evaluation controls visual leakage, and that small omni-modal models can gain substantial capability through carefully staged post-training and self-distilled omni-query supervision. We release the OmniClean evaluation data to support leakage-aware omni-modal evaluation.
\end{abstract}

\section{Introduction} \label{sec:introduction}

Recent omni-modal language models aim to provide a unified interface for understanding audio, visual inputs, and language~\cite{xu2025qwen25omnireport,xu2025qwen3omnireport,yang2025humanomniv2understandingomnimodalreasoning,liu2025nexuso}. However, strong benchmark performance does not necessarily imply genuine omni-modal integration. In many audio-visual-language tasks, visual evidence and the question can already be sufficient to recover the answer, allowing models to score well without using audio. As a result, raw benchmark gains may reflect visual shortcut exploitation rather than improved omni-modal understanding~\cite{agrawal2018overcoming,zhou2026dailyomniaudiovisualreasoningtemporal,yang2025humanomniv2understandingomnimodalreasoning}.

We address this issue by constructing \textbf{OmniClean}\footnote{\url{https://huggingface.co/datasets/che111/OmniClean}}, a visually debiased evaluation view over nine existing omni benchmarks. We audit each query with visual-only probing, remove visually solvable queries, and retain full subsets only for benchmark-specific exception cases where filtering is undefined or would make score comparisons unstable. This protocol audits 16,968 queries and yields 8,551 retained queries. OmniClean is therefore an operational evaluation view: it reduces visual shortcuts under a fixed protocol rather than proving that the retained queries are causally audio-dependent in every possible setting.

Using OmniClean, we study \textbf{OmniBoost}, a staged post-training recipe based on Qwen2.5-Omni-3B~\cite{xu2025qwen25omnireport}. The study asks whether strengthening the constituent bi-modal abilities, namely vision-language and audio-language understanding, is enough for omni-modal understanding, or whether explicit omni-modal data and optimization signals are needed. To answer this, we compare a balanced mixed bi-modal supervised fine-tuning (SFT) control following common instruction-tuning practice~\cite{10.5555/3600270.3602281,wang2022selfinstruct,liu2023visual}, mixed-modality reinforcement learning with verifiable rewards (RLVR)~\cite{shao2024deepseekmath,deepseekai2025,yu2025dapoopensource}, and SFT on self-distilled data~\cite{hinton2015distilling,wu2025sdrt}. For self-distillation, we construct synthetic omni-modal queries without relying on a stronger external omni-modal teacher. Instead, an entity-based procedure derives spatial and temporal relations from LLaVA-Video seed clips~\cite{zhang2024videoinstructiontuningsynthetic}, Step-Audio-R1 audio captions~\cite{tian2025stepaudior1technicalreport}, Qwen3-VL video captions~\cite{bai2025qwen3vltechnicalreport}, and gpt-oss-120b entity scaffolds~\cite{openai2025gptoss}, then converts them into hard-matchable audio-visual-text questions before filtering model-generated reasoning traces. The results show that balanced bi-modal SFT alone gives limited and uneven transfer, whereas the first broad improvement appears only after training with explicit omni-modal data. The competitive gains come from the staged post-training recipe and the synthetic-query construction: after SFT on self-distilled data, the 3B model becomes comparable to larger open-source references and slightly exceeds Qwen3-Omni-30B-A3B-Instruct~\cite{xu2025qwen3omnireport} under both OmniClean aggregate summaries, without distilling answers from a stronger omni-modal teacher.

The rest of the paper is organized as follows. Section~2 reviews omni-modal models, audio-visual-language evaluation, and post-training. Section~3 presents the visual-leakage audit and OmniClean construction. Section~4 reports the OmniBoost staged post-training study, and Section~5 summarizes the main findings and limitations.

\section{Background and Related Work}

\subsection{Omni-modal LLMs}
Recent multimodal systems have expanded beyond vision-language or audio-language settings toward \emph{omni-modal} interfaces that can consume text, images, video, and audio within a single model. Representative recent systems include Qwen2.5-Omni~\cite{xu2025qwen25omnireport}, Qwen3-Omni~\cite{xu2025qwen3omnireport}, HumanOmniV2~\cite{yang2025humanomniv2understandingomnimodalreasoning}, NEXUS-O~\cite{liu2025nexuso}, and Nemotron 3 Nano Omni~\cite{deshmukh2026nemotron3nanoomni}. Modern vision-language models such as Qwen3-VL~\cite{bai2025qwen3vltechnicalreport}, InternVL3.5~\cite{wang2025internvl35}, and Molmo2~\cite{clark2026molmo2} continue to advance visual understanding, while audio-language models such as Step-Audio~\cite{huang2025stepaudio}, Step-Audio 2~\cite{wu2025stepaudio2}, Step-Audio-R1~\cite{tian2025stepaudior1technicalreport}, and Step-Audio-R1.5~\cite{zhang2026stepaudior15} focus on audio-centric instruction following and reasoning. Omni-modal language models extend these lines by integrating audio, visual inputs, and language in a single interface.

However, access to multiple modalities does not guarantee omni-modal integration: visually dominant evidence can make some queries answerable without audio, causing evaluations to overestimate omni-modal capability. Related bias and shortcut effects have long been discussed in multimodal evaluation~\cite{agrawal2018overcoming} and are increasingly acknowledged in recent omni-modal work~\cite{zhou2026dailyomniaudiovisualreasoningtemporal,yang2025humanomniv2understandingomnimodalreasoning}. This motivates evaluation protocols that can separate genuine omni-modal use from cases where performance is largely explained by unimodal competence.

\subsection{Audio-Visual-Language Evaluation}
Recent audio-visual-language benchmarks aim to measure whether a model can jointly understand audio-visual events and answer language queries grounded in omni-modal evidence, as in Daily-Omni~\cite{zhou2026dailyomniaudiovisualreasoningtemporal}, WorldSense~\cite{hong2026worldsenseevaluatingrealworldomnimodal}, OmniBench~\cite{li2025omnibenchfutureuniversalomnilanguage}, IntentBench~\cite{yang2025humanomniv2understandingomnimodalreasoning}, AV-Odyssey~\cite{gong2024avodysseybenchmultimodalllms}, Video-Holmes~\cite{cheng2025videoholmesmllmthinklike}, UNO-Bench~\cite{chen2025unobenchunifiedbenchmarkexploring}, CG-AV-Counting~\cite{lu2025avreasonerimprovingbenchmarkingcluegrounded}, and OmniVideoBench~\cite{li2026omnivideobenchaudiovisualunderstandingevaluation}. Collectively, these benchmarks cover temporal alignment, intent and social reasoning, counting, complex video reasoning, and open-world audio-visual QA, and many provide verifiable targets such as multiple-choice answers or numeric outputs. However, verifiability alone does not prevent modality leakage: some queries remain solvable from visual content and the question alone, causing benchmark scores to conflate omni-modal understanding with visual shortcut exploitation.

\subsection{Post-Training for Multimodal Models}
Post-training improves instruction following and reasoning in multimodal models through supervised fine-tuning (SFT) on curated or synthetic data~\cite{10.5555/3600270.3602281,wang2022selfinstruct,liu2023visual}, reinforcement-learning-style optimization with verifiable or task-aligned rewards~\cite{Schulman2017ProximalPO,shao2024deepseekmath,deepseekai2025,yu2025dapoopensource}, and distillation or self-distillation~\cite{hinton2015distilling,wu2025sdrt}. For omni-modal models, the unresolved question is whether vision-language and audio-language competence can simply compose, or whether explicit omni-modal signals are required; recent multimodal RL work~\cite{huang2025vision,wang2025vl,wan2025srpo,jiaqi2025think,deshmukh2026nemotron3nanoomni} suggests that targeted optimization can improve reasoning, motivating our staged study under a visually debiased evaluation view.

\section{Probing Visual Leakage and Constructing a Cleaned Evaluation View}

This section revisits existing omni benchmarks through the lens of \emph{visual leakage}. The central question is whether an ostensibly audio-visual-language query can still be answered from visual input and the question alone. We therefore probe existing benchmarks with visual-only probing, compare the original and cleaned score views where that comparison is defined, and construct a visually debiased evaluation view with benchmark-specific full-retention exceptions under our protocol.

\subsection{Visual-Only Probing and a Cleaned Evaluation View}

Our audit is operational. For each evaluation query, we keep the image or video together with the text question, withhold the audio input, and test whether a strong model can still recover the correct verifiable answer. If a query passes verification under this visual-only setting, we mark it as visually answerable and exclude it from the cleaned evaluation view; otherwise we retain it. This criterion reduces visual shortcuts under our protocol rather than proving exclusive audio dependence.

\paragraph{Evaluation and verification protocol.} For score reporting, we follow the official evaluation setting and answer format of each source benchmark~\cite{zhou2026dailyomniaudiovisualreasoningtemporal,hong2026worldsenseevaluatingrealworldomnimodal,li2025omnibenchfutureuniversalomnilanguage,yang2025humanomniv2understandingomnimodalreasoning,gong2024avodysseybenchmultimodalllms,cheng2025videoholmesmllmthinklike,chen2025unobenchunifiedbenchmarkexploring,lu2025avreasonerimprovingbenchmarkingcluegrounded,li2026omnivideobenchaudiovisualunderstandingevaluation}. For video inputs, we sample frames at 2 fps. If a video exceeds 60 seconds, we uniformly sample 120 frames over the full clip; otherwise we use all frames sampled at 2 fps under the same 120-frame budget. Each video frame is resized so that the shorter edge is 448 pixels while preserving the original aspect ratio. Image inputs are passed directly unless the shorter edge exceeds 768 pixels, in which case the image is resized to a 768-pixel shorter edge with the aspect ratio preserved. The model receives the benchmark-native media and the original question and options. We do not add an extra system prompt, modality hint, or task-specific chain-of-thought instruction. For visual-only probing, we sample 16 rollouts per query with temperature set to 1.0 and a maximum generation length of 8192 tokens. Reported score evaluations use the same input preprocessing and verifier but are run separately from the pass@16 probing procedure.

The answer space is verifiable: most queries are multiple-choice questions with letter or option-text answers, and the remaining evaluated queries have numeric targets. We therefore use benchmark-aware normalization followed by hard matching against the official gold answer. For multiple-choice questions, we accept either the final option letter or the normalized option text after removing leading option markers such as ``A.'', ``(A)'', or ``A:''. For numeric answers, we canonicalize signs, commas, and decimal notation and compare the resulting numeric value, using an official benchmark tolerance only when the source benchmark defines one.

\paragraph{Cleaning protocol.} Unless otherwise noted, visual-only cleaning is performed with \textbf{Qwen3-VL-30B-A3B-Thinking}~\cite{bai2025qwen3vltechnicalreport}. For each query, we provide only the visual input together with the original text question, generate 16 visual-only rollouts using the input construction above, and remove the query if at least one rollout is verified as correct. This pass@16 rule is used only to construct the cleaned split and to produce the visual-only probing histograms; reported model scores on the original or filtered views are fresh evaluations under the official benchmark settings on the corresponding query set. This distinction is why a model used in the cleaning probe can still obtain a non-zero score when evaluated again on the retained filtered subset: the retained set is not a proof of impossibility under every prompt or decode, but an operational set of queries not solved under the fixed visual-only screening run. We apply the same rule to all applicable benchmarks in this section for diagnostic probing. The final evaluation construction has two exceptions. For AV-Odyssey~\cite{gong2024avodysseybenchmultimodalllms}, we do not define a filtered subset under this protocol because some answer options themselves contain audio input that a pure VL model cannot directly consume; accordingly, all score-based comparisons retain the full evaluation subset. For CG-AV-Counting~\cite{lu2025avreasonerimprovingbenchmarkingcluegrounded}, we still run visual-only probing for diagnosis, but we do not report a filtered evaluation subset from this 376-query subset because further exclusion would substantially reduce evaluation stability.

\begin{figure}[ht]
    \centering
    \begin{subfigure}[b]{0.24\textwidth}
        \centering
        \includegraphics[width=\textwidth]{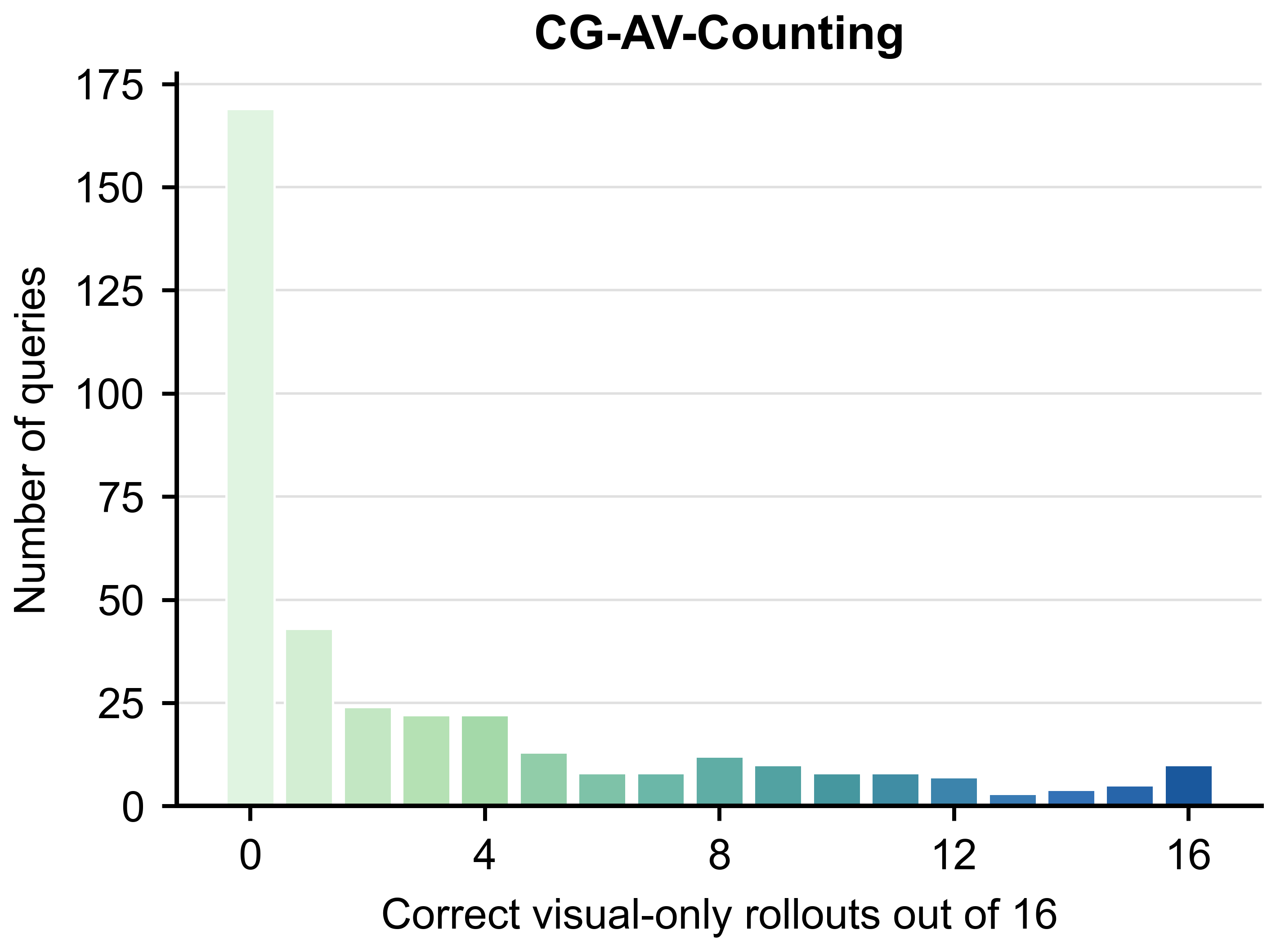}
        \caption{CG-AV-Counting~\cite{lu2025avreasonerimprovingbenchmarkingcluegrounded}}
    \end{subfigure}
    \hfill
    \begin{subfigure}[b]{0.24\textwidth}
        \centering
        \includegraphics[width=\textwidth]{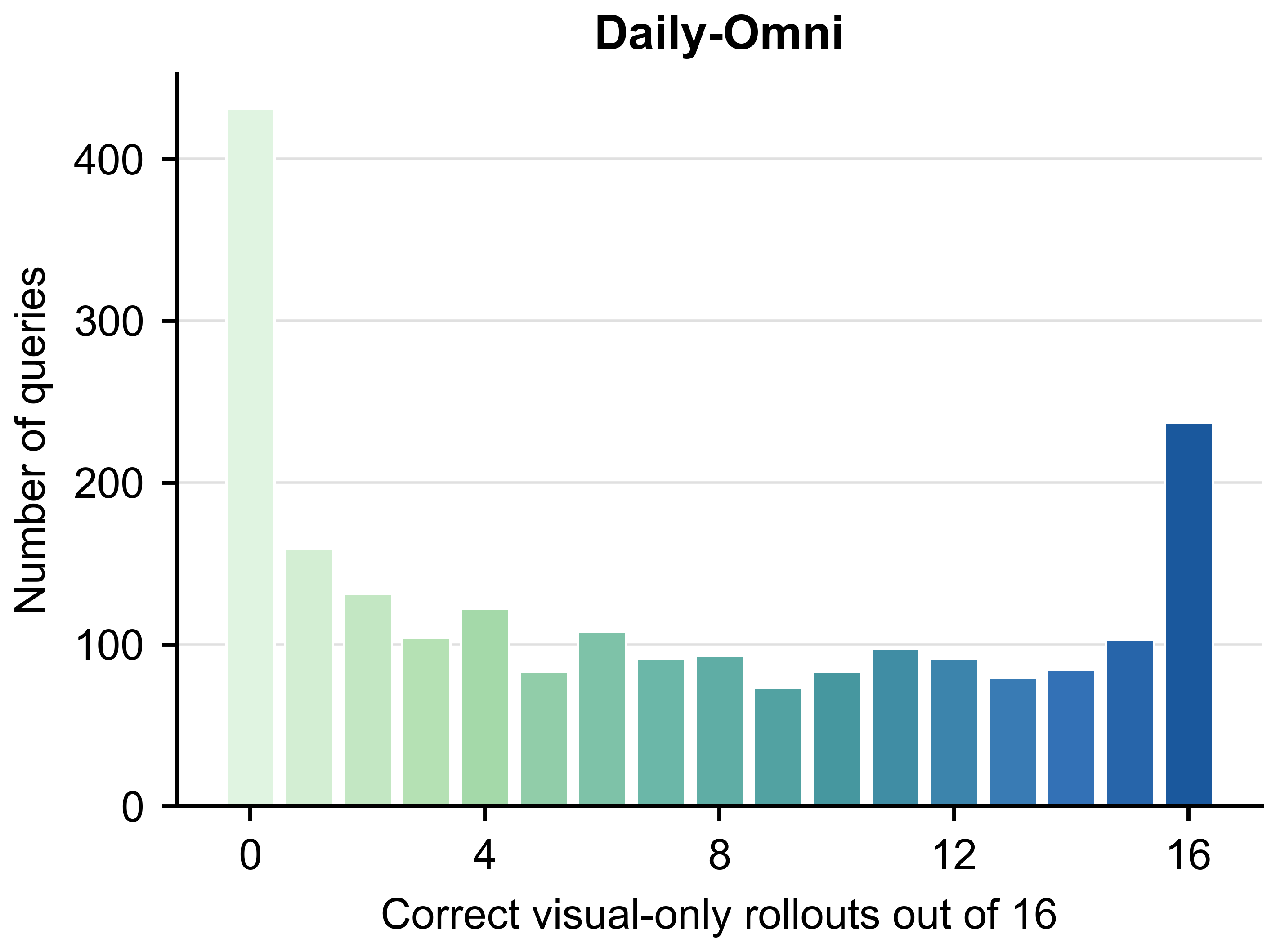}
        \caption{Daily-Omni~\cite{zhou2026dailyomniaudiovisualreasoningtemporal}}
    \end{subfigure}
    \hfill
    \begin{subfigure}[b]{0.24\textwidth}
        \centering
        \includegraphics[width=\textwidth]{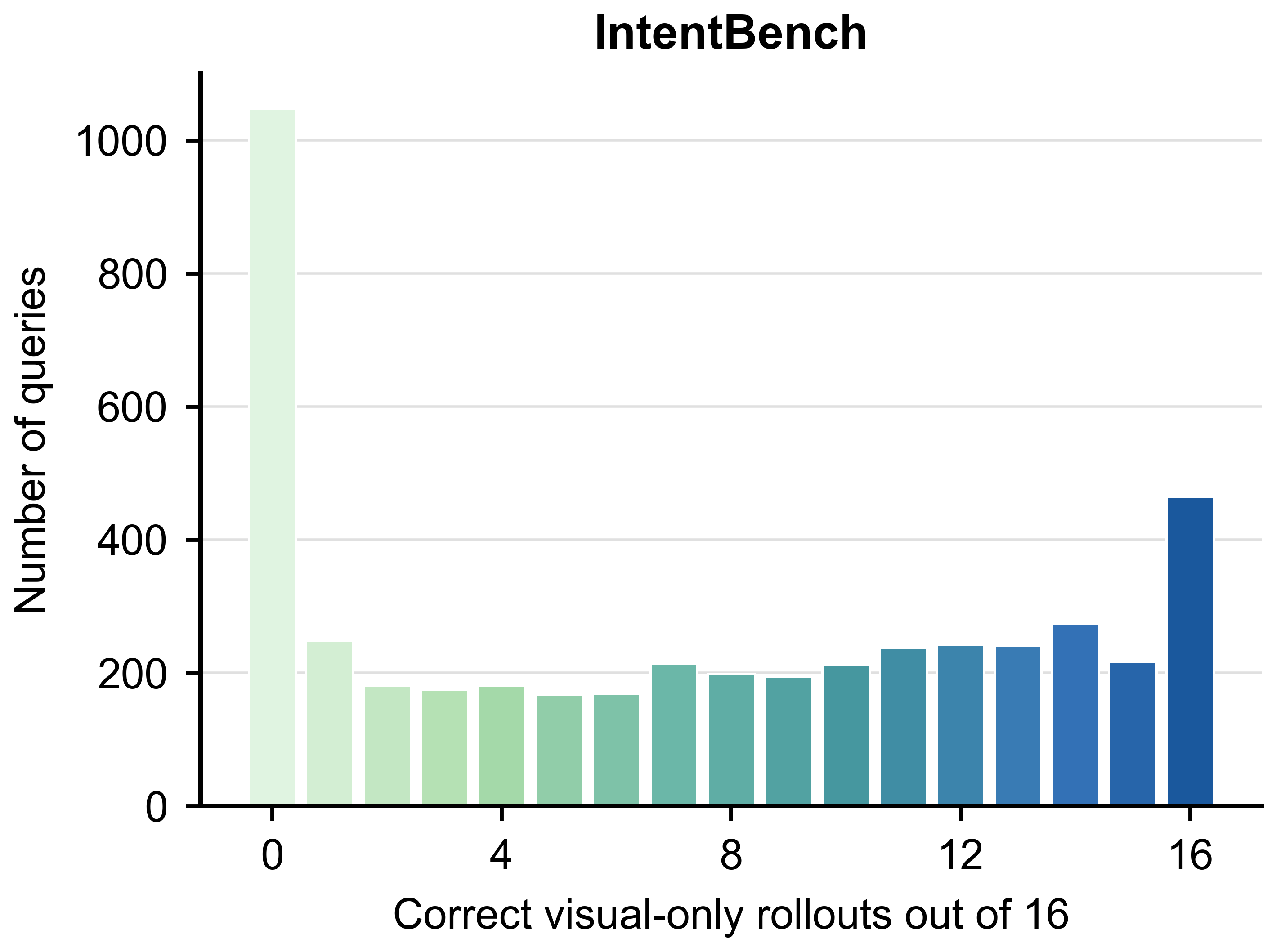}
        \caption{IntentBench~\cite{yang2025humanomniv2understandingomnimodalreasoning}}
    \end{subfigure}
    \hfill
    \begin{subfigure}[b]{0.24\textwidth}
        \centering
        \includegraphics[width=\textwidth]{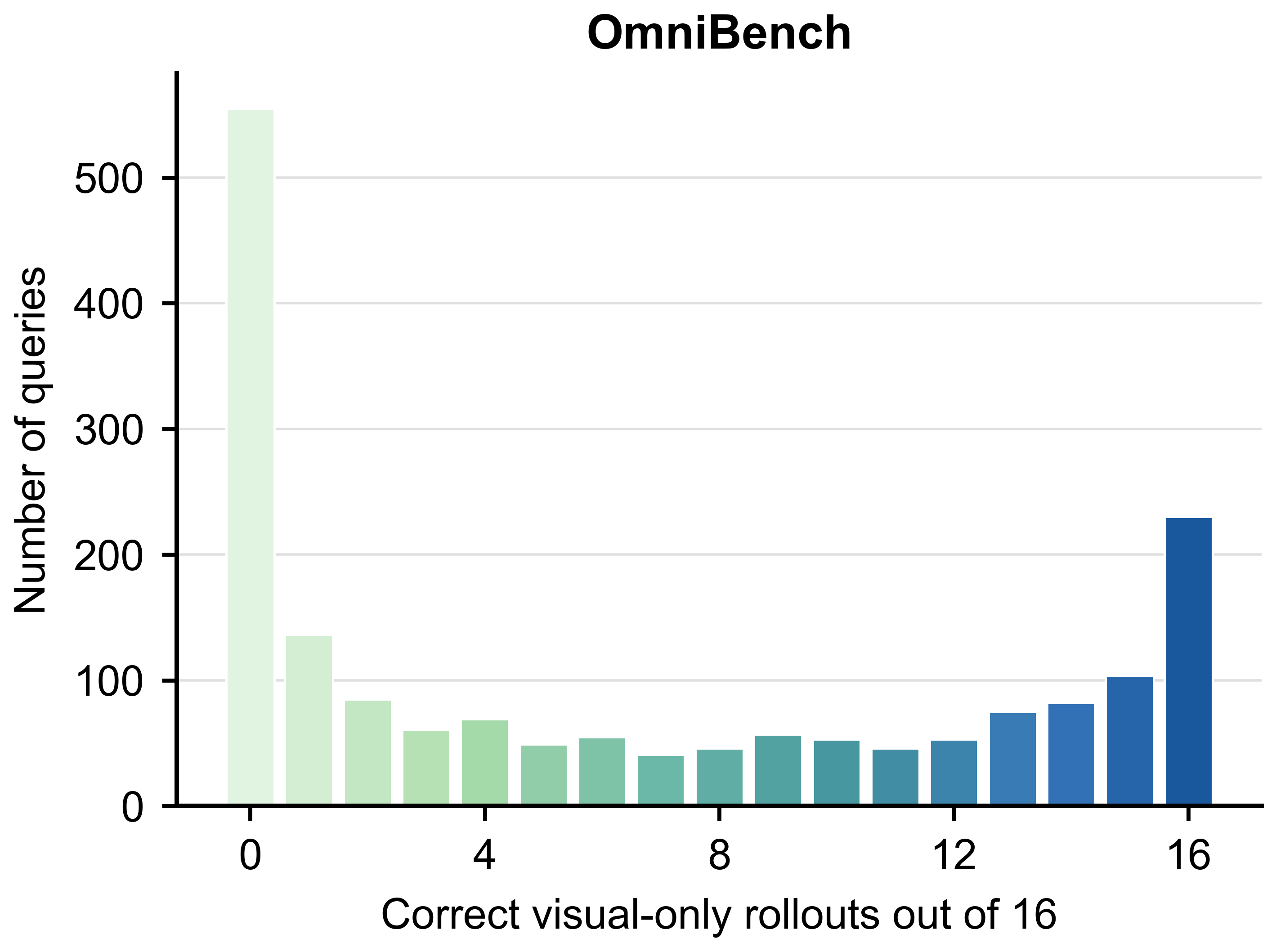}
        \caption{OmniBench~\cite{li2025omnibenchfutureuniversalomnilanguage}}
    \end{subfigure}
    
    \vspace{1em}
    
    \hfill
    \begin{subfigure}[b]{0.24\textwidth}
        \centering
        \includegraphics[width=\textwidth]{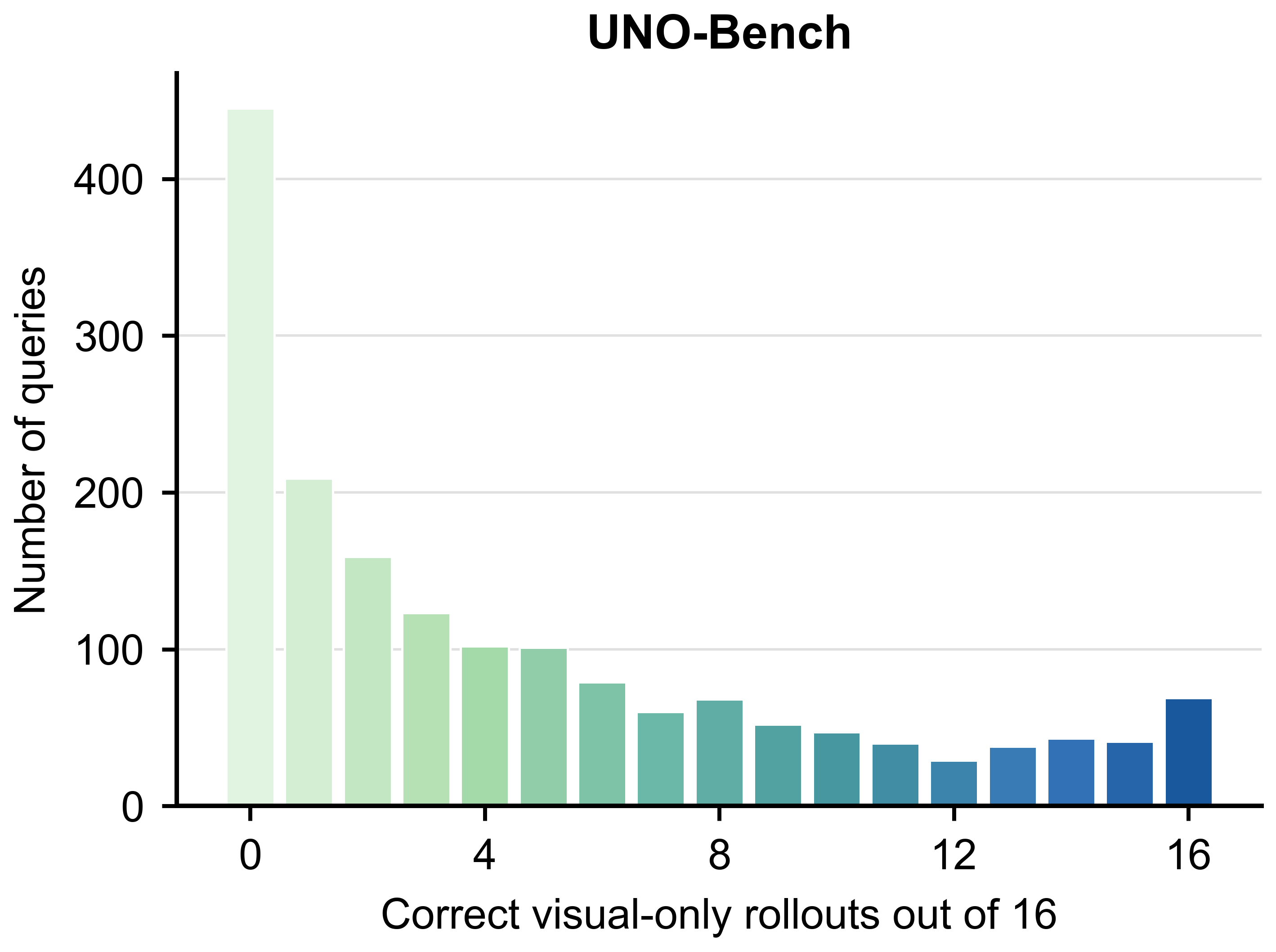}
        \caption{UNO-Bench~\cite{chen2025unobenchunifiedbenchmarkexploring}}
    \end{subfigure}
    \hfill
    \begin{subfigure}[b]{0.24\textwidth}
        \centering
        \includegraphics[width=\textwidth]{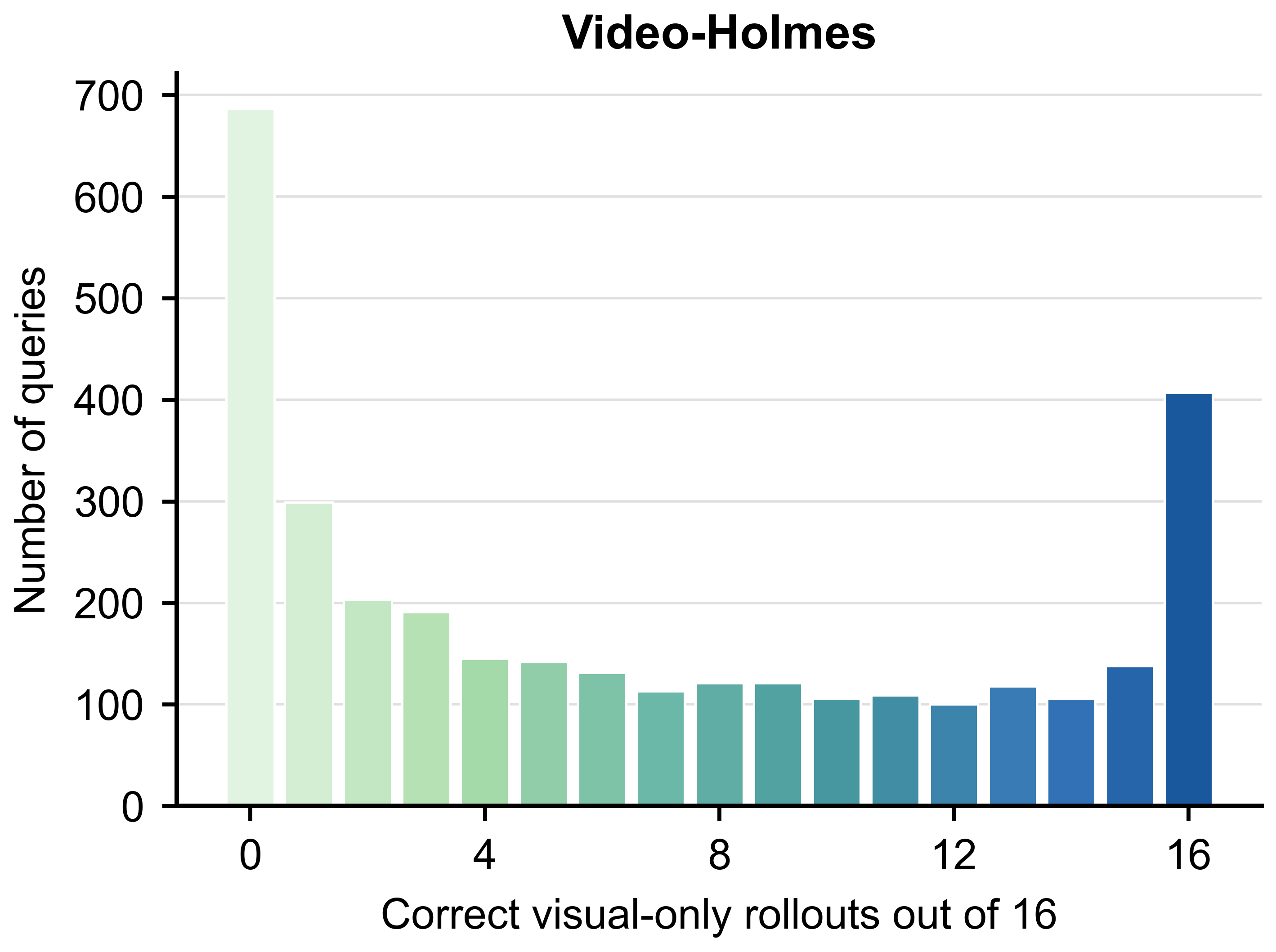}
        \caption{Video-Holmes~\cite{cheng2025videoholmesmllmthinklike}}
    \end{subfigure}
    \hfill
    \begin{subfigure}[b]{0.24\textwidth}
        \centering
        \includegraphics[width=\textwidth]{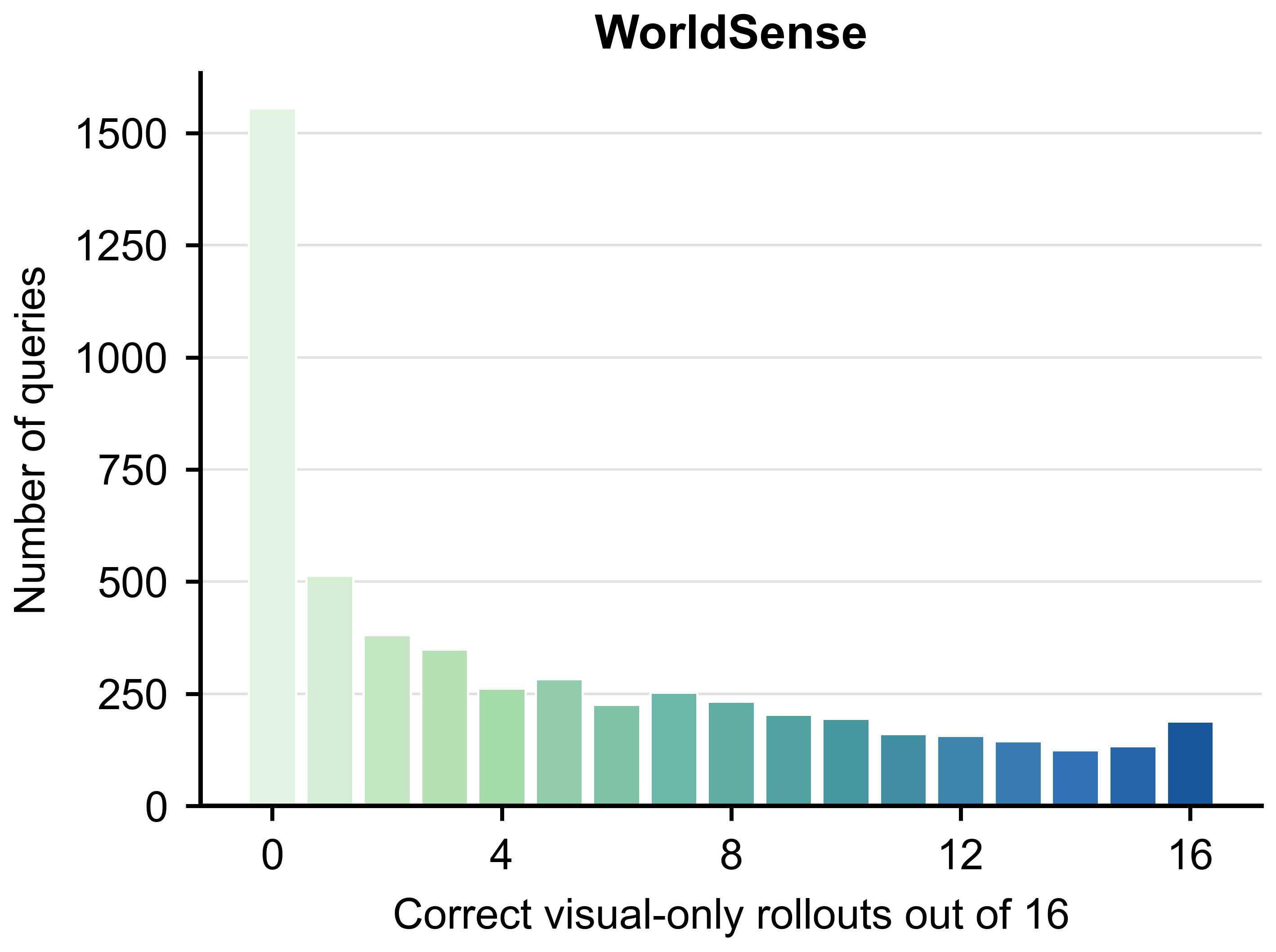}
        \caption{WorldSense~\cite{hong2026worldsenseevaluatingrealworldomnimodal}}
    \end{subfigure}
    \hfill
    \begin{subfigure}[b]{0.24\textwidth}
        \centering
        \includegraphics[width=\textwidth]{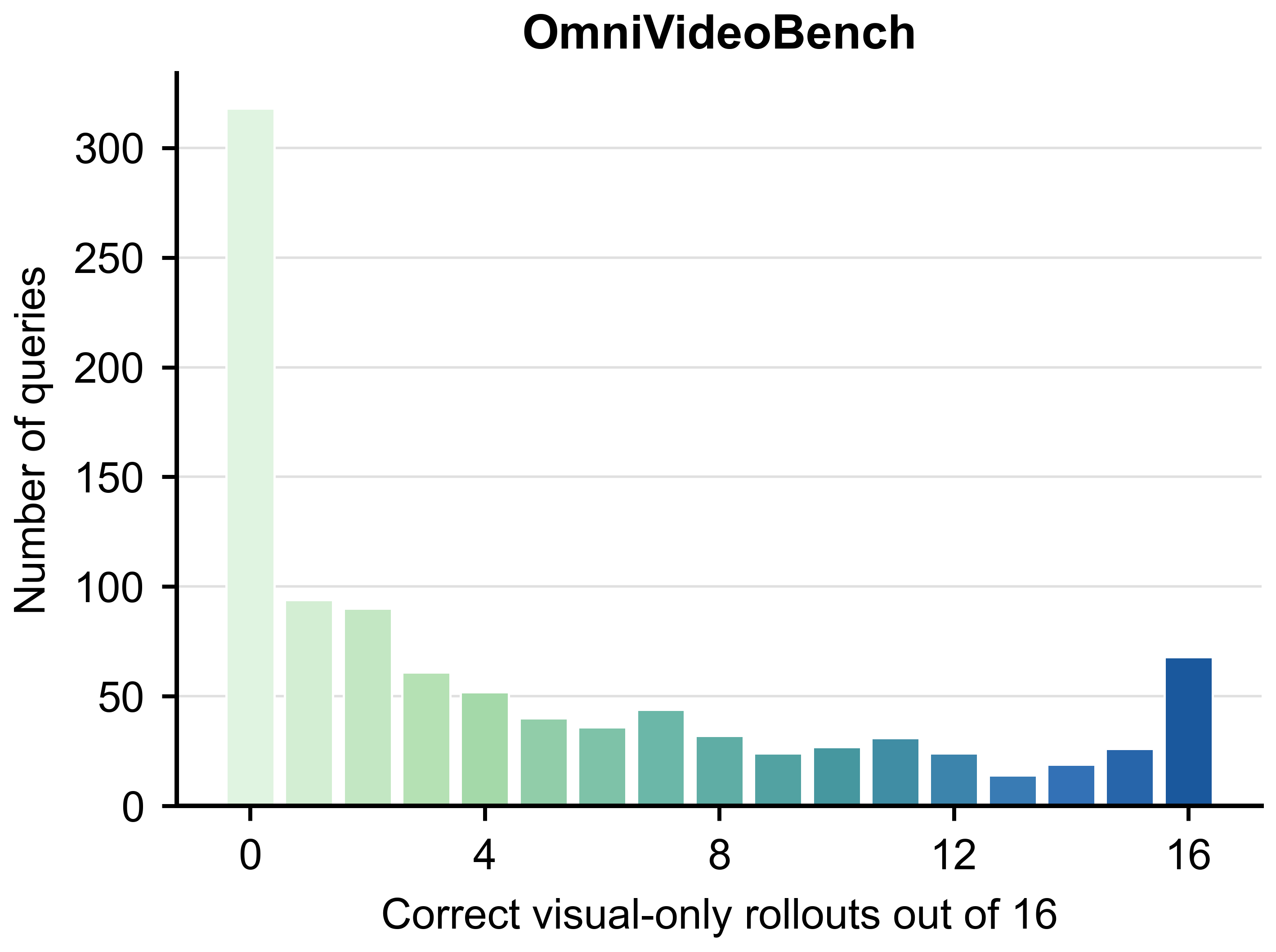}
        \caption{OmniVideoBench~\cite{li2026omnivideobenchaudiovisualunderstandingevaluation}}
    \end{subfigure}
    \hfill
    
        \caption{Visual-only probing outcomes across applicable benchmarks. Histograms show the number of correct visual-only rollouts per query; mass near zero indicates fewer visually solvable queries, while mass at higher counts indicates visual leakage. AV-Odyssey~\cite{gong2024avodysseybenchmultimodalllms} is omitted because its answer options contain audio-bearing input.}
    \label{fig:v_only_correctness}
\end{figure}

Figure~\ref{fig:v_only_correctness} shows large benchmark-level variation in visual-only solvability: Daily-Omni~\cite{zhou2026dailyomniaudiovisualreasoningtemporal} and OmniBench~\cite{li2025omnibenchfutureuniversalomnilanguage} contain a substantial share of queries solved by visual-only rollouts, whereas Video-Holmes~\cite{cheng2025videoholmesmllmthinklike} retains a larger visually unsolved core. AV-Odyssey~\cite{gong2024avodysseybenchmultimodalllms} is omitted because its answer options can contain audio input, making this visual-only screening protocol undefined. The histogram therefore motivates query-level cleaning rather than relying only on aggregate benchmark scores.

\begin{table}[ht]
    \centering
\caption{Original and filtered scores across audited benchmarks when a filtered-score view is defined. ``Filtered'' denotes official-protocol evaluation on the retained query set, with red deltas relative to original scores; the pass@16 visual-only rule is used only to construct the split. Reference columns use the Qwen3-VL~\cite{bai2025qwen3vltechnicalreport}, Qwen2.5-Omni~\cite{xu2025qwen25omnireport}, and Qwen3-Omni~\cite{xu2025qwen3omnireport} model families. Benchmark sources are cited in the benchmark notes below.}
    \label{tab:sketch}
    \newcommand{\rotmodel}[2]{\rotatebox[origin=l]{28}{\shortstack[l]{\scriptsize\textbf{#1}\\[-1pt]\scriptsize\textbf{#2}}}}
    \newcommand{\scoredrop}[1]{{\scriptsize\textcolor{red}{(-#1)}}}
    \resizebox{\textwidth}{!}{
    \begin{tabular}{l|l|cccccc}
        \toprule[1.2pt]
        \textbf{Benchmark}
        & \textbf{Split}
        & \rotmodel{Qwen3-VL}{30B-A3B-Instruct}
        & \rotmodel{Qwen3-VL}{30B-A3B-Thinking}
        & \rotmodel{Qwen2.5-Omni}{3B}
        & \rotmodel{Qwen2.5-Omni}{7B}
        & \rotmodel{Qwen3-Omni}{30B-A3B-Instruct}
        & \rotmodel{Qwen3-Omni}{30B-A3B-Thinking} \\
        \midrule
        \multirow{2}{*}{\textbf{Daily-Omni}}
            & \textbf{Original} & 53.64 & 54.57 & 46.86 & 51.51 & 57.65 & 70.65 \\
            & \textbf{Filtered} & 1.05 \scoredrop{52.59} & 1.77 \scoredrop{52.80} & 27.53 \scoredrop{19.33} & 31.78 \scoredrop{19.73} & 31.22 \scoredrop{26.43} & 42.62 \scoredrop{28.03} \\
        \midrule
        \multirow{2}{*}{\textbf{IntentBench}}
            & \textbf{Original} & 55.93 & 57.40 & 44.06 & 51.06 & 57.36 & 65.38 \\
            & \textbf{Filtered} & 20.71 \scoredrop{35.22} & 20.24 \scoredrop{37.16} & 29.57 \scoredrop{14.49} & 31.61 \scoredrop{19.45} & 32.46 \scoredrop{24.90} & 36.42 \scoredrop{28.96} \\
        \midrule
        \multirow{2}{*}{\textbf{Video-Holmes}}
            & \textbf{Original} & 39.37 & 44.66 & 28.65 & 31.37 & 42.44 & 53.63 \\
            & \textbf{Filtered} & 33.45 \scoredrop{5.92} & 32.97 \scoredrop{11.69} & 24.36 \scoredrop{4.29} & 27.37 \scoredrop{4.00} & 40.94 \scoredrop{1.50} & 46.33 \scoredrop{7.30} \\
        \midrule
        \multirow{2}{*}{\textbf{WorldSense}}
            & \textbf{Original} & 40.63 & 40.60 & 37.17 & 40.28 & 43.83 & 51.27 \\
            & \textbf{Filtered} & 2.23 \scoredrop{38.40} & 1.91 \scoredrop{38.69} & 24.91 \scoredrop{12.26} & 24.25 \scoredrop{16.03} & 23.79 \scoredrop{20.04} & 27.70 \scoredrop{23.57} \\
        \midrule
        \multirow{2}{*}{\textbf{OmniBench}}
            & \textbf{Original} & 35.03 & 35.65 & 37.59 & 43.10 & 48.29 & 54.87 \\
            & \textbf{Filtered} & 3.88 \scoredrop{31.15} & 3.22 \scoredrop{32.43} & 27.14 \scoredrop{10.45} & 32.12 \scoredrop{10.98} & 32.97 \scoredrop{15.32} & 32.15 \scoredrop{22.72} \\
        \midrule
        \multirow{2}{*}{\textbf{UNO-Bench}}
            & \textbf{Original} & 31.22 & 32.27 & 27.25 & 30.46 & 41.11 & 52.17 \\
            & \textbf{Filtered} & 4.07 \scoredrop{27.15} & 4.24 \scoredrop{28.03} & 21.41 \scoredrop{5.84} & 24.84 \scoredrop{5.62} & 29.17 \scoredrop{11.94} & 37.55 \scoredrop{14.62} \\
        \midrule
        \multirow{2}{*}{\textbf{CG-AV-Counting}}
            & \textbf{Original} & 15.66 & 19.65 & 12.73 & 15.13 & 18.57 & 20.28 \\
            & \textbf{Filtered} & -- & -- & -- & -- & -- & -- \\
        \midrule
        \multirow{2}{*}{\textbf{OmniVideoBench}}
            & \textbf{Original} & 30.82 & 29.29 & 35.80 & 33.70 & 38.50 & 39.02 \\
            & \textbf{Filtered} & 10.24 \scoredrop{20.58} & 2.12 \scoredrop{27.17} & 27.67 \scoredrop{8.13} & 29.25 \scoredrop{4.45} & 32.90 \scoredrop{5.60} & 31.27 \scoredrop{7.75} \\
        \midrule
        \multirow{2}{*}{\textbf{AV-Odyssey}}
            & \textbf{Original} & -- & -- & 29.00 & 30.16 & 32.61 & 40.02 \\
            & \textbf{Filtered} & -- & -- & -- & -- & -- & -- \\
        \bottomrule[1.2pt]
    \end{tabular}}
    
    \vspace{0.35em}
    
    {\footnotesize \textit{Note.} For AV-Odyssey~\cite{gong2024avodysseybenchmultimodalllms}, the VL-only columns and all ``Filtered'' entries are omitted because a visual-only filtered subset is not defined: some answer options require audio input, which pure VL models cannot accept. For CG-AV-Counting~\cite{lu2025avreasonerimprovingbenchmarkingcluegrounded}, ``Filtered'' entries are also omitted, but for a different reason: the visual-only probing is used only to diagnose visual solvability, while all score-based comparisons retain the full evaluation subset. Because filtered-score views are not uniformly defined across the audited suite, this table is used for benchmark-level leakage diagnosis rather than aggregate model ranking. Red deltas after the filtered scores denote absolute score drops relative to the corresponding original-score row.}
\end{table}

\begin{figure}[ht]
    \centering
    \includegraphics[width=0.72\linewidth]{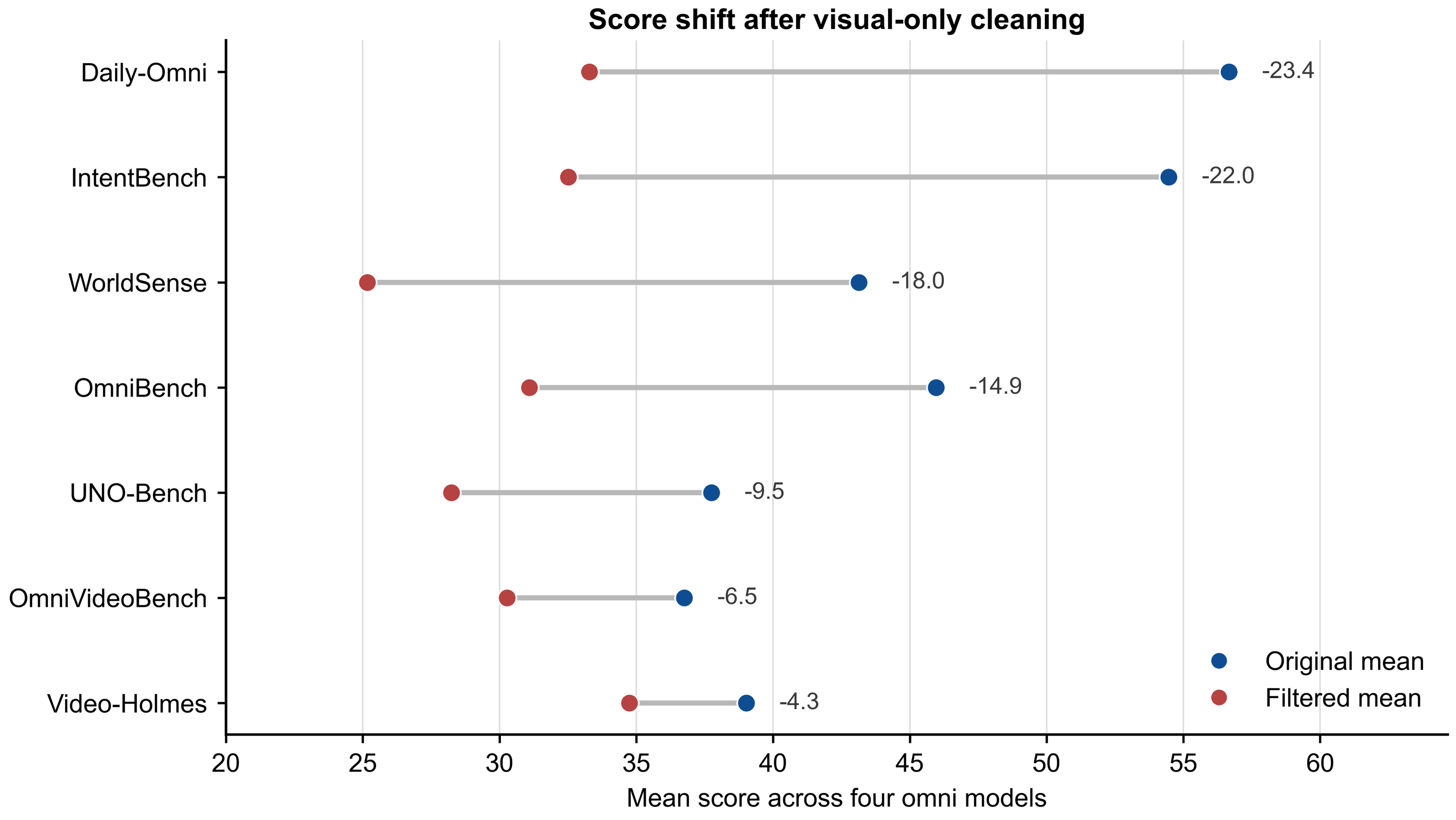}
    \caption{Score distributions before and after query-level cleaning for benchmarks with both original and filtered score views. The separation between the original and cleaned distributions summarizes how strongly visually answerable queries affect reported omni-modal performance.}
    \label{fig:omni_performance_boxplot}
\end{figure}

Figure~\ref{fig:omni_performance_boxplot} and Table~\ref{tab:sketch} together show that visual leakage is highly uneven across benchmarks. Daily-Omni~\cite{zhou2026dailyomniaudiovisualreasoningtemporal} and OmniBench~\cite{li2025omnibenchfutureuniversalomnilanguage} lose a large fraction of apparent omni performance after filtering, whereas Video-Holmes~\cite{cheng2025videoholmesmllmthinklike} preserves a larger retained core. We intentionally do not report a macro or query-weighted average in Table~\ref{tab:sketch}: the table is a leakage diagnostic, and filtered-score views are not uniformly defined for the full audited suite. AV-Odyssey~\cite{gong2024avodysseybenchmultimodalllms} and CG-AV-Counting~\cite{lu2025avreasonerimprovingbenchmarkingcluegrounded} are excluded from these filtered-score summaries for different reasons: AV-Odyssey lacks a defined visual-only filtered subset because its answer options contain audio-bearing input, while CG-AV-Counting is probed diagnostically but retained fully for score stability.

For reference, the benchmark notes below distinguish three quantities when needed: the original scale reported by the source paper, the pre-cleaning query count used in our audited evaluation view, and the retained query count after applying our protocol. The audited suite spans image-grounded, video-grounded, counting, intent, and open-ended QA settings:
\begin{itemize}
    \item \textbf{Daily-Omni}~\cite{zhou2026dailyomniaudiovisualreasoningtemporal}: a multiple-choice audio-visual QA benchmark for temporally aligned reasoning in daily scenarios, with 684 real-world videos and 1,197 questions across six task families. We audit all 1,197 queries in this study and retain 237 queries after visual-only cleaning.
    \item \textbf{IntentBench}~\cite{yang2025humanomniv2understandingomnimodalreasoning}: a benchmark for reasoning about human intention, emotion, and deception from jointly grounded audio-visual context, with 633 videos and 2,689 questions. We audit all 2,689 queries and retain 660 after cleaning.
    \item \textbf{Video-Holmes}~\cite{cheng2025videoholmesmllmthinklike}: a complex video reasoning benchmark built from suspense short films that requires models to connect distributed clues over time, with 270 videos and 1,837 question-answer pairs across seven tasks. We audit all 1,837 queries and retain 885 after cleaning.
    \item \textbf{WorldSense}~\cite{hong2026worldsenseevaluatingrealworldomnimodal}: a real-world omnimodal video benchmark emphasizing strong audio-video coupling, containing 1,662 synchronized videos and 3,172 multiple-choice QA pairs across 26 tasks. We audit all 3,172 queries and retain 875 after cleaning.
    \item \textbf{OmniBench}~\cite{li2025omnibenchfutureuniversalomnilanguage}: a human-annotated tri-modal benchmark for joint reasoning over visual, acoustic, and textual inputs, containing 1,142 questions designed to require integrated evidence across modalities. We audit all 1,142 queries and retain 417 after cleaning.
    \item \textbf{UNO-Bench}~\cite{chen2025unobenchunifiedbenchmarkexploring}: a unified benchmark spanning 44 task types and five modality combinations; the original release contains 1,250 omni-modal samples and 2,480 uni-modal samples. Our evaluation uses only its 1,000-query multiple-choice UNOBench-MC subset as the pre-cleaning audited view, from which 228 queries are retained after cleaning.
    \item \textbf{AV-Odyssey}~\cite{gong2024avodysseybenchmultimodalllms}: a large-scale multiple-choice benchmark for audio-visual understanding with interleaved text, visual, and audio evidence, covering 4,555 problems across 26 tasks and 10 domains. We audit all 4,555 problems and retain the full evaluation subset in the final evaluation because its answer options contain audio-bearing content that a pure VL model cannot directly accept, so a visual-only filtered subset is not defined under our protocol.
    \item \textbf{CG-AV-Counting}~\cite{lu2025avreasonerimprovingbenchmarkingcluegrounded}: a clue-grounded audio-visual counting benchmark over long videos, with 497 videos, 1,027 multimodal questions, and 5,845 manually annotated clues. In our experiments, we use a 376-query subset selected from examples annotated by the dataset as requiring both audio and video, excluding audio-only or video-only cases. We run the same visual-only probing analysis on this subset for diagnosis, but we do not construct or report a filtered-subset benchmark from it. The benchmark is already highly challenging under the probe, and further exclusion would substantially shrink the effective subset and reduce evaluation stability, so all score-based comparisons retain the full evaluation subset.
    \item \textbf{OmniVideoBench}~\cite{li2026omnivideobenchaudiovisualunderstandingevaluation}: an audio-visual video understanding benchmark with manually verified QA, containing 628 videos and 1,000 question-answer pairs across 13 question types. We audit all 1,000 queries and retain 318 after cleaning.
\end{itemize}
Overall, across the selected evaluation suite studied here, the filtering unit is the query rather than the underlying media item. We audit 16,968 queries before cleaning and retain 8,551 queries after cleaning or full retention under the rules above. We release this final cleaned evaluation view as \textbf{OmniClean}, a visually debiased evaluation dataset over the same nine audited omni benchmarks.

\subsection{Correlation Shifts After Cleaning}
After the leakage diagnosis, we use correlation and regression analyses as supporting diagnostics for how cleaning changes benchmark meaning. The analysis asks whether cleaned scores become less tied to uni-modal vision or audio strength and more reflective of intended omni-modal evidence use. These correlations are descriptive, computed over the four open-source omni models with available original, filtered, vision, and audio reference scores; AV-Odyssey and CG-AV-Counting are omitted because they do not have reported filtered-score views.

\begin{figure}[t]
    \centering
    \includegraphics[width=0.94\linewidth]{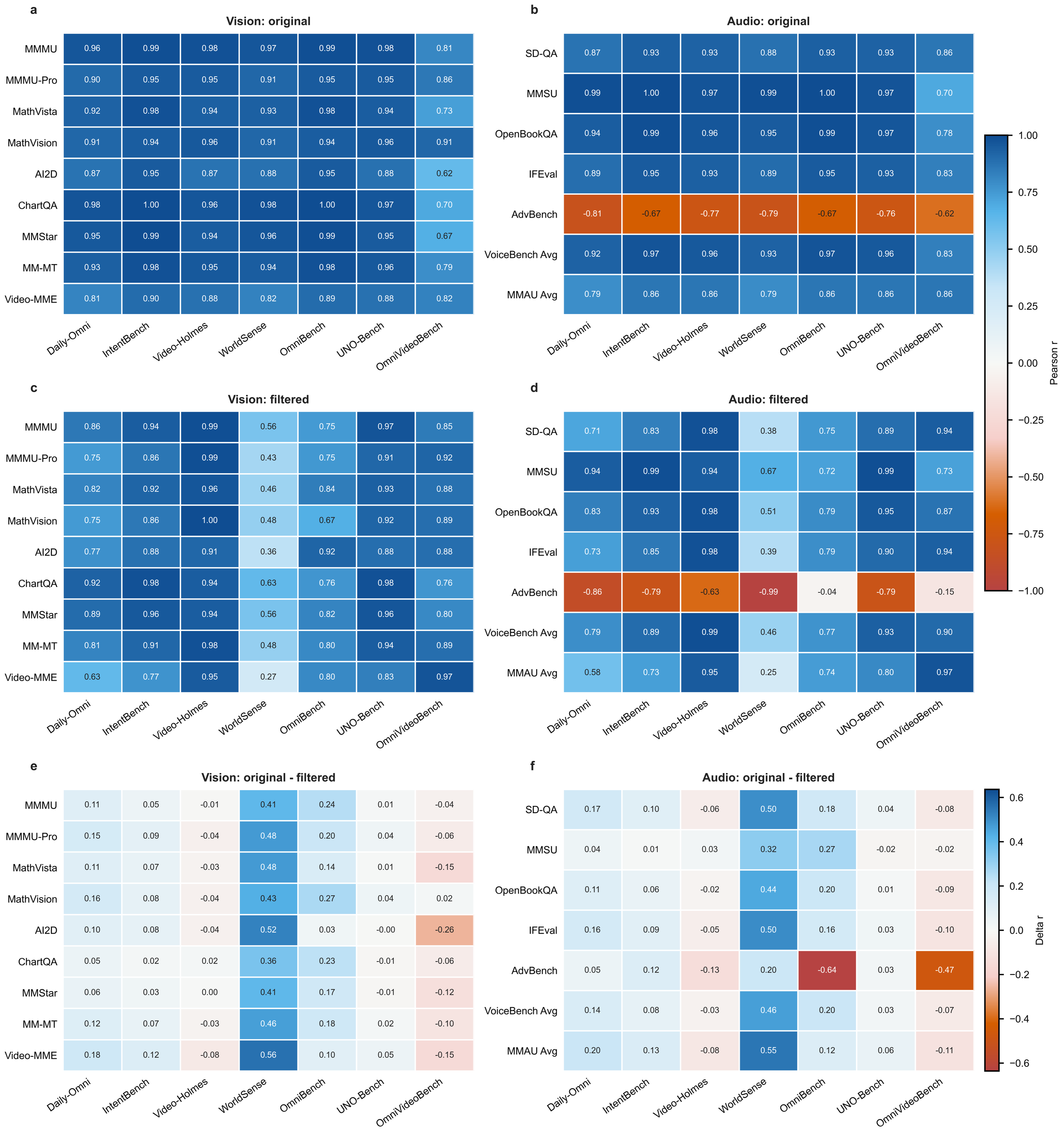}
    \caption{Correlation shifts after cleaning. Columns correspond to vision and audio uni-modal reference views. Rows show correlations on the original query set, correlations on the cleaned query set, and the gap $\Delta r=\text{Original}-\text{Filtered}$. Positive gap values mean that the cleaned score is less correlated with that uni-modal reference, while negative values indicate a stronger correlation after cleaning.}
    \label{fig:correlation_shifts_cleaned}
\end{figure}

The correlation-shift diagnostic in Figure~\ref{fig:correlation_shifts_cleaned} shows that cleaning changes what several benchmarks track. WorldSense~\cite{hong2026worldsenseevaluatingrealworldomnimodal} exhibits the largest correlation shift, with both vision- and audio-side correlations dropping substantially after filtering. Daily-Omni~\cite{zhou2026dailyomniaudiovisualreasoningtemporal}, IntentBench~\cite{yang2025humanomniv2understandingomnimodalreasoning}, OmniBench~\cite{li2025omnibenchfutureuniversalomnilanguage}, and UNO-Bench~\cite{chen2025unobenchunifiedbenchmarkexploring} also become less dominated by uni-modal reference strength, whereas Video-Holmes~\cite{cheng2025videoholmesmllmthinklike} and OmniVideoBench~\cite{li2026omnivideobenchaudiovisualunderstandingevaluation} show smaller or mixed shifts. Thus, filtering changes benchmark meaning in a dataset-dependent way rather than uniformly lowering all uni-modal correlations.

\subsection{How Uni-modal Capabilities Predict Omni Scores}

We next test whether omni scores can be predicted from uni-modal reference strength alone. On the original views, visual strength is often a strong predictor, matching the leakage diagnosis. After filtering, this relationship weakens or shifts for several benchmarks, indicating that cleaned scores are less uniformly explained by broad uni-modal competence. The complete benchmark-by-benchmark regression gallery is reported in Appendix~\ref{appendix:full_unimodal_regression}, and the exact source pools are listed in Appendix~\ref{appendix:regression_source_pools}.

\subsection{Toward a Cleaned Evaluation View}
The audit suggests that omni evaluation should report visual-shortcut sensitivity explicitly and compare original and cleaned views where defined. We release \textbf{OmniClean} as a cleaned evaluation view over nine existing benchmarks, preserving verifiable answer formats while reducing visual shortcuts under our visual-only probing protocol. Section~4 uses this view to evaluate post-training signals under a less shortcut-sensitive setting.

\section{OmniBoost: A Staged Post-Training Study}

This section presents \textbf{OmniBoost}, our staged post-training study on the cleaned evaluation view introduced in Section~3. We use Qwen2.5-Omni-3B~\cite{xu2025qwen25omnireport} as the base model. OmniBoost includes a strong mixed bi-modal SFT control following supervised fine-tuning practice~\cite{10.5555/3600270.3602281,wang2022selfinstruct,liu2023visual}, a mixed-modality RLVR stage~\cite{shao2024deepseekmath,deepseekai2025,yu2025dapoopensource} that delivers broad cleaned-view gains, and a self-distillation SFT stage~\cite{hinton2015distilling,wu2025sdrt}; an additional fixed-setup ablation shows that filtered synthetic self-distillation data can directly improve the base model.

\subsection{Staged Post-Training Study Design}

We organize OmniBoost around two linked post-training questions: whether balanced bi-modal supervision is sufficient for cleaned omni-modal gains, and whether explicit omni-modal data plus later self-distillation can further improve model capability. To test these questions, we use three completed stages under a shared initialization lineage: mixed bi-modal SFT, mixed-modality RLVR, and self-distillation SFT.

\subsubsection{Data Construction Across the Staged Study}
The study draws on three corresponding training pools:
\begin{enumerate}
    \item \textbf{Balanced Mixed Bi-modal SFT Pool:} A four-way mixture of audio-text, image-text, video-text, and pure-text supervision, with each source sampled to 1B output tokens.
    \item \textbf{Mixed-Modality RLVR Pool:} A curated mixed-modality optimization set spanning text-only, image-text, video-text, audio-image-text, and audio-video-text queries, optimized with DAPO~\cite{yu2025dapoopensource}.
    \item \textbf{Synthetic Audio-Visual-Text Pool for Self-Distillation SFT:} A synthetic query set constructed from LLaVA-Video seed videos~\cite{zhang2024videoinstructiontuningsynthetic}, segment-level audio and video captions, and entity-relation records; the Stage~2 RLVR checkpoint then generates candidate reasoning traces that are filtered before self-distillation SFT~\cite{wu2025sdrt}. The main self-distillation SFT result additionally adjusts data ratios on top of this shared pipeline.
\end{enumerate}

\subsubsection{Stage 1: Mixed Bi-modal SFT}
We first establish a deliberately strong control baseline using mixed bi-modal supervision only, without adding explicit omni-modal instruction data in this stage. The purpose is to test whether large-scale aggregation of dual-modal competence can already transfer to the cleaned omni evaluation.

Our mixed bi-modal SFT baseline starts from \textbf{Qwen2.5-Omni-3B}~\cite{xu2025qwen25omnireport} and is output-token balanced across four sources: audio-text (1B output tokens), image-text (1B), video-text (1B), and pure text (1B). The audio-text, image-text, and pure-text portions are drawn from internal datasets. The video-text portion combines four open-source video corpora: Video-R1-data~\cite{feng2025videor1}, VideoAuto-R1-Data~\cite{liu2026videoautor1}, ShareGPT4Video~\cite{chen2024sharegpt4video}, and LLaVA-Video-178K~\cite{zhang2024videoinstructiontuningsynthetic}. Because these corpora partially overlap, we deduplicate exact matches at the video-query level while retaining multiple distinct queries for the same video when appropriate. We then rewrite the video CoTs with Qwen2.5-VL-235B~\cite{bai2025qwen25vltechnicalreport}, add dense full-video captions derived from 30-second segments, and discard examples that the 235B model still cannot answer correctly. This construction gives each source the same output-token budget so that the comparison focuses on modality composition rather than simple data imbalance.

\paragraph{Training setup.} We train this SFT stage for 1 epoch with a global batch size of 64. Training examples are packed into 64K-token sequences using modality-agnostic packing, so text-only, audio-text, image-text, and video-text examples can share packed sequences. Data from the four sources are mixed by direct shuffling, and we do not impose additional batch-level balancing beyond the dataset-level output-token budget described above. This stage therefore serves as a controlled first-round composition study rather than an exhaustive hyperparameter search.

\subsubsection{Stage 2: Mixed-Modality RLVR}
Starting from the 1-epoch mixed bi-modal SFT checkpoint, we apply RLVR to refine the model's reasoning behavior. In this stage, the main design goal is to optimize queries that explicitly require omni-grounded reasoning while replaying visual and textual queries for robustness. The later self-distillation analyses then use this RLVR checkpoint as a common starting point for controlled follow-up comparisons.

\begin{figure}[!htbp]
    \centering
    \includegraphics[width=0.92\linewidth]{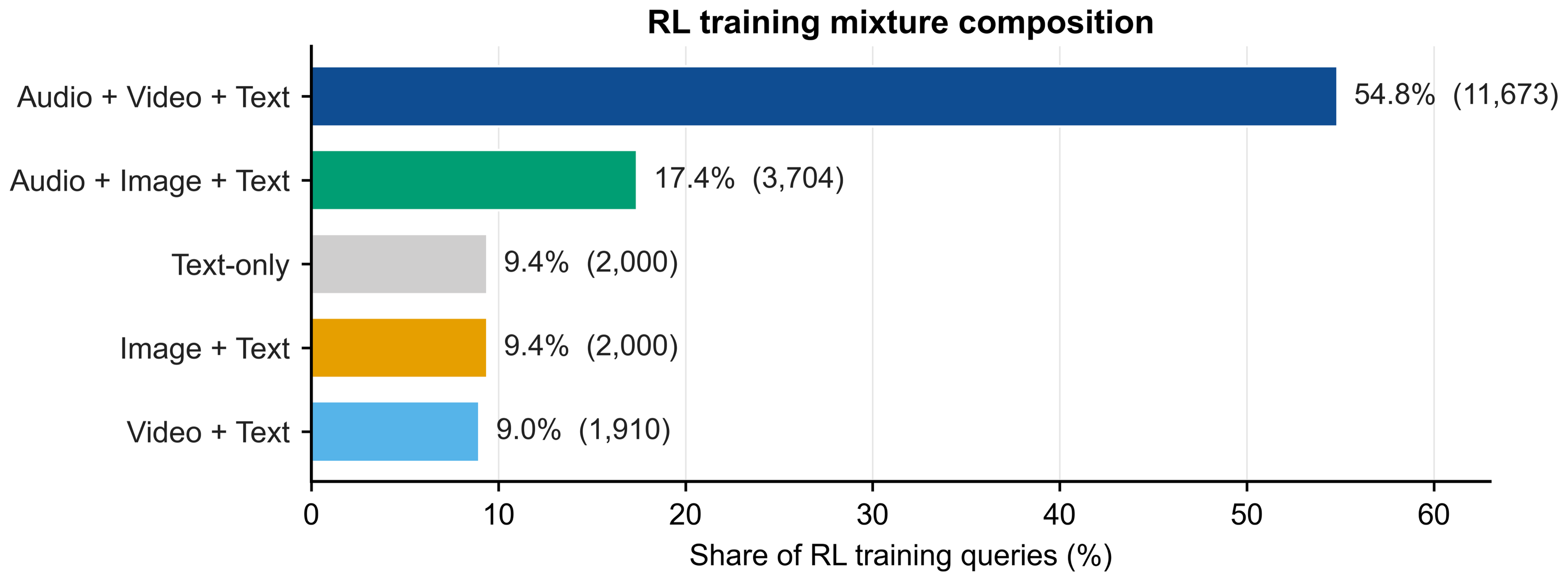}
    \caption{Modality composition of the RLVR training mixture. Ranked horizontal bars show both percentage share and query count; every category includes a text question.}
    \label{fig:rl_mixture_composition}
\end{figure}

For this stage, we construct curated training queries that explicitly span pure text, image-text, video-text, audio-image-text, and audio-video-text settings. The goal is to make the optimization target depend on omni-modal evidence integration instead of merely rewarding strong performance on any single modality. We also keep visual and textual replay queries in the mixture so that broader capability is not discarded while the optimization target shifts toward omni-grounded correctness. The resulting RLVR mixture contains 54.8\% audio-video-text queries, 17.4\% audio-image-text queries, 9.0\% video-text queries, 9.4\% image-text queries, and 9.4\% text-only queries; all five categories include an explicit text question, as visualized in Figure~\ref{fig:rl_mixture_composition}. Within this staged study, RLVR is the first training component that yields substantial benchmark-level macro gains on the cleaned evaluation view.

\paragraph{Training setup.} We use DAPO~\cite{yu2025dapoopensource} as the RLVR algorithm in this stage, without adding a KL penalty term. Each query is rolled out 16 times. At the optimization level, each update selects 32 queries, and each query is paired with 16 rollouts, giving a total batch size of 512 trajectories per update. We set the maximum generation length to 4K, the sampling temperature to 1.0, and the learning rate to $1\times10^{-6}$. The reported result comes from a 1200-step RLVR run initialized from the 1-epoch mixed bi-modal SFT checkpoint.

\paragraph{Reward schedule.} Our reward design follows a simple two-stage schedule. During the first 500 steps, we combine a format reward and an accuracy reward with weights 0.8 and 0.2, respectively, to stabilize structured generation early in training. After step 500, once the response format becomes much more stable, we reduce the format reward weight to 0.1 and increase the accuracy-ratio reward weight to 0.9 so that optimization focuses more directly on correct grounded answers.

\subsubsection{Stage 3: Self-Distillation SFT with Filtered Synthetic Queries}
Starting from the RLVR 1200-step checkpoint, we build a self-distillation SFT stage around synthetic audio-visual-text question-answer pairs with verifiable answer formats~\cite{wu2025sdrt}. This stage uses dense audio captions and dense video descriptions as construction-time evidence to synthesize richer queries, then samples multiple rollouts per query and filters them through a quality-control pipeline before distillation SFT. At a high level, we construct synthetic queries that jointly expose audio, video, and textual task instructions while keeping the generated answers in hard-matchable forms such as option indices, option text, numbers, or short phrases. This synthetic pool is broader than the original RLVR set and is designed to amplify useful reasoning patterns at scale without giving up verifiability.

\begin{figure}[!htbp]
    \centering
    \includegraphics[width=\linewidth]{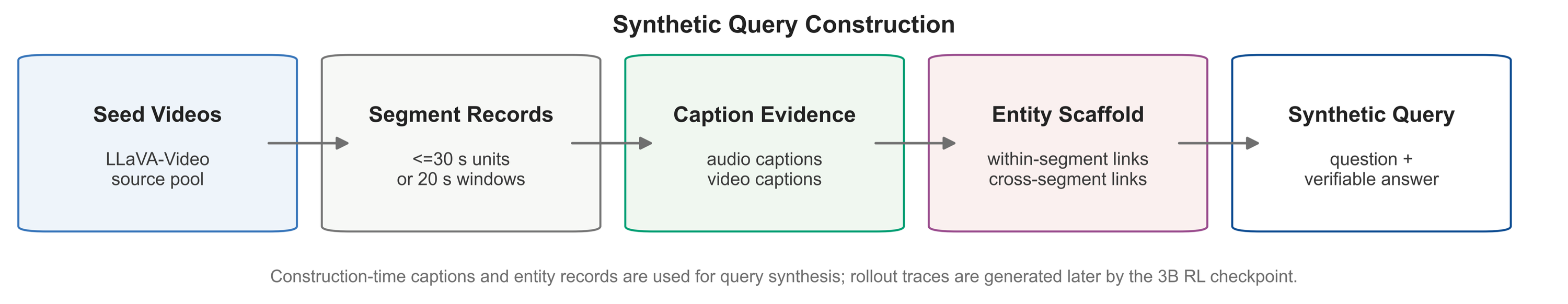}
    \caption{Synthetic Query construction before rollout filtering. LLaVA-Video seeds~\cite{zhang2024videoinstructiontuningsynthetic} are segmented, captioned with Step-Audio-R1~\cite{tian2025stepaudior1technicalreport} and Qwen3-VL-235B-A22B~\cite{bai2025qwen3vltechnicalreport}, and summarized into caption/entity records. gpt-oss-120b~\cite{openai2025gptoss} then composes hard-matchable Synthetic Query pairs from the captions, scaffold, and answer-format constraints. Appendix~\ref{appendix:synthetic_query_graphic_description} provides the detailed graphic description.}
    \label{fig:synthetic_query_construction}
\end{figure}

\paragraph{Synthetic Query construction.} We select seed videos from the LLaVA-Video source pool~\cite{zhang2024videoinstructiontuningsynthetic} and use caption/entity records, rather than raw media, as construction-time input for question synthesis. Videos of at most 30 seconds are treated as single units; longer videos are annotated in 20-second windows, with a final remainder shorter than 10 seconds merged into the preceding segment and a remainder longer than 10 seconds kept separately. Each segment receives an audio caption from \textbf{Step-Audio-R1}~\cite{tian2025stepaudior1technicalreport} and a detailed video caption from \textbf{Qwen3-VL-235B-A22B}~\cite{bai2025qwen3vltechnicalreport}. From these segment-level records, we extract recurring entities and ask \textbf{gpt-oss-120b}~\cite{openai2025gptoss} to organize them into a lightweight entity-relation graph over within-segment relations and cross-segment temporal links. This graph is a relation scaffold for Synthetic Query construction, not a formal claim of complete spatio-temporal annotation: a 20-second segment can itself contain temporal dynamics as well as spatial or event co-occurrence. Conditioned on the captions, entity graph, and desired answer format, the language model composes candidate question-answer pairs. The distillation traces are not produced at this step; they are generated later by the 3B RLVR checkpoint during rollout sampling. The final training instance pairs the original media input with the Synthetic Query and the generated hard-match answer target, while malformed question-answer pairs, answer leakage, inconsistent options, and caption-entity mismatches are removed before rollout generation. Figure~\ref{fig:synthetic_query_construction} summarizes this construction pipeline.

\paragraph{Main self-distillation setup.} The main Stage~3 result reported in Table~\ref{tab:sec4_stage_results} is a ratio-adjusted self-distillation SFT run initialized from the RLVR 1200-step checkpoint. It should be read as the third stage in the OmniBoost trajectory: Qwen2.5-Omni-3B $\rightarrow$ mixed bi-modal SFT $\rightarrow$ mixed-modality RLVR $\rightarrow$ self-distillation SFT. The answer traces used for this self-distillation stage are generated by the same 3B lineage after RLVR and filtered with generated hard-match answer targets, rather than distilled from a stronger external omni-model answer teacher. The synthetic pool is formed by combining retained video-centric and audio-centric Synthetic Query data after the F1--F3 quality-control passes described below; the final Stage~3 checkpoint uses a ratio-adjusted mixture selected from the same candidate pool family instead of treating any single pass-retained dataset as the final recipe. During self-distillation rollout generation, each Synthetic Query is sampled 8 times from the RLVR checkpoint to provide candidate reasoning traces before filtering. The SFT objective is standard next-token supervised learning on the retained trace and answer text, using the original media input paired with the Synthetic Query. This main Stage~3 result is distinct from the fixed-setup filtering ablation in Table~\ref{tab:self_distill_stage_results}, which intentionally restarts from the same Qwen2.5-Omni-3B baseline, trains each pass-specific dataset for 60 steps with packed 64K-token sequences and a learning rate of $1\times10^{-5}$, and is used only to isolate the value of different filtered synthetic datasets.

\paragraph{Self-distillation filtering passes.} To avoid overloading the word ``stage,'' we refer to the quality-control steps inside self-distillation as filtering passes \textbf{F1--F3}. These passes are applied progressively: F2 is run on the data retained after F1, and F3 is run on the data retained after F2. For each Synthetic Query, we start from the 1200-step RLVR checkpoint and generate 8 rollouts. \textbf{F1} filters by rollout difficulty profile: we remove queries whose 8 rollouts are all wrong, and we also remove queries that are solved too uniformly, i.e., with 7/8 or 8/8 correct rollouts. \textbf{F2} removes traces with clear perception defects or malformed outputs from the F1-retained pool. In practice, this pass drops rollouts and queries whose reasoning explicitly shows missing perception (e.g., the model states that it cannot hear or cannot see the relevant evidence), and it also removes generations that contain abnormal media tokens such as \texttt{<audio>} or \texttt{<image>} inside the output. \textbf{F3} then enforces consistency between the reasoning trace and the final answer on the F2-retained pool. We keep only rollouts whose reasoning and answer agree with the generated hard-match answer target after normalization; if the reasoning arrives at the target option but the final answer tag points to a different option, we rewrite the answer tag to match the choice implied by the reasoning. The distilled SFT stage then reuses the retained traces to strengthen the reasoning patterns that RLVR first makes available.

\subsection{Main Staged Results on OmniClean}

Unless otherwise stated, the staged post-training variants in this section follow a single lineage from \textbf{Qwen2.5-Omni-3B}~\cite{xu2025qwen25omnireport}. Mixed bi-modal SFT starts from this base model, RLVR starts from the 1-epoch mixed bi-modal SFT checkpoint, and self-distillation SFT starts from the RLVR 1200-step checkpoint. The staged comparisons below should therefore be read as controlled post-training variants within the same model lineage rather than as separate model families.

We use the benchmark-level macro average as the primary summary because the benchmark families differ substantially in size and task design. Query-weighted averages are reported as a complementary view of the retained-query mixture, not as the basis for the main stage-ordering claim.

\begin{table}[ht]
    \centering
    \caption{Scores on OmniClean for open-source omni references and the three OmniBoost stages within the same Qwen2.5-Omni-3B lineage~\cite{xu2025qwen25omnireport}; Qwen3-Omni references are cited to the Qwen3-Omni report~\cite{xu2025qwen3omnireport}. Stage~3 is the ratio-adjusted self-distillation SFT run initialized from the RLVR 1200-step checkpoint; it is not the same fixed-setup ablation as Table~\ref{tab:self_distill_stage_results}. Macro averages weight benchmarks equally, while query-weighted averages weight retained query counts; averages use retained counts and unrounded scores. AV-Odyssey~\cite{gong2024avodysseybenchmultimodalllms} and CG-AV-Counting~\cite{lu2025avreasonerimprovingbenchmarkingcluegrounded} are retained as full subsets under the exception rules described in Section~3, where all benchmark sources are cited.}
    \label{tab:sec4_stage_results}
    \resizebox{\textwidth}{!}{
    \begin{tabular}{l|cccc|ccc}
        \toprule[1.2pt]
        \textbf{Benchmark}
        & \shortstack[c]{\textbf{Qwen2.5-Omni} \\ \textbf{3B}}
        & \shortstack[c]{\textbf{Qwen2.5-Omni} \\ \textbf{7B}}
        & \shortstack[c]{\textbf{Qwen3-Omni} \\ \textbf{30B-A3B-Instruct}}
        & \shortstack[c]{\textbf{Qwen3-Omni} \\ \textbf{30B-A3B-Thinking}}
        & \shortstack[c]{\textbf{Stage 1} \\ \textbf{Mixed Bi-modal SFT}}
        & \shortstack[c]{\textbf{Stage 2} \\ \textbf{Mixed-Modality RLVR}}
        & \shortstack[c]{\textbf{Stage 3} \\ \textbf{Self-Distillation SFT}} \\
        \midrule
        \textbf{Daily-Omni}      & 27.53 & 31.78 & 31.22 & 42.62 & 27.43 & 38.05 & 38.82 \\
        \textbf{IntentBench}     & 29.57 & 31.61 & 32.46 & 36.42 & 30.15 & 36.46 & 37.03 \\
        \textbf{Video-Holmes}    & 24.36 & 27.37 & 40.94 & 46.33 & 31.53 & 47.07 & 44.46 \\
        \textbf{WorldSense}      & 24.91 & 24.25 & 23.79 & 27.70 & 24.11 & 27.53 & 24.71 \\
        \textbf{OmniBench}       & 27.14 & 32.12 & 32.97 & 32.15 & 32.13 & 43.24 & 40.29 \\
        \textbf{UNO-Bench}       & 21.41 & 24.84 & 29.17 & 37.55 & 23.68 & 21.97 & 23.35 \\
        \textbf{CG-AV-Counting}  & 12.73 & 15.13 & 18.57 & 20.28 & 16.22 & 19.65 & 16.49 \\
        \textbf{OmniVideoBench}  & 27.67 & 29.25 & 32.90 & 31.27 & 25.16 & 21.00 & 22.33 \\
        \textbf{AV-Odyssey}      & 29.00 & 30.16 & 32.61 & 40.02 & 28.00 & 27.87 & 31.80 \\
        \midrule
        \textbf{Macro Avg.}      & 24.92 & 27.39 & 30.51 & 34.93 & 26.49 & 31.43 & 31.03 \\
        \textbf{Query-Weighted Avg.} & 27.05 & 28.68 & 31.84 & 37.56 & 27.58 & 30.74 & 32.15 \\
        \bottomrule[1.2pt]
    \end{tabular}}
\end{table}

Table~\ref{tab:sec4_stage_results} summarizes the three reported OmniBoost stages within the same model lineage; Appendix~\ref{appendix:stage_delta_heatmap} visualizes the same cleaned-view stage deltas relative to Qwen2.5-Omni-3B~\cite{xu2025qwen25omnireport}. Under the benchmark-level macro average, performance improves from 26.49 for \textbf{Stage 1: Mixed Bi-modal SFT} to 31.43 for \textbf{Stage 2: Mixed-Modality RLVR}, while \textbf{Stage 3: Self-Distillation SFT} reaches 31.03. This pattern shows that balanced mixed bi-modal SFT alone is not sufficient for consistent omni gains on OmniClean: its macro improvement is modest and benchmark-level changes remain uneven. This makes Stage~2 the strongest OmniBoost stage under the benchmark-family summary and supports the need for explicit omni-modal data rather than only broader dual-modal coverage. Relative to Stage~2, Stage~3 improves AV-Odyssey~\cite{gong2024avodysseybenchmultimodalllms}, Daily-Omni~\cite{zhou2026dailyomniaudiovisualreasoningtemporal}, IntentBench~\cite{yang2025humanomniv2understandingomnimodalreasoning}, OmniVideoBench~\cite{li2026omnivideobenchaudiovisualunderstandingevaluation}, and UNO-Bench~\cite{chen2025unobenchunifiedbenchmarkexploring}, but Stage~2 remains stronger on CG-AV-Counting~\cite{lu2025avreasonerimprovingbenchmarkingcluegrounded}, OmniBench~\cite{li2025omnibenchfutureuniversalomnilanguage}, Video-Holmes~\cite{cheng2025videoholmesmllmthinklike}, and WorldSense~\cite{hong2026worldsenseevaluatingrealworldomnimodal}. The query-weighted average changes the ordering: Stage~3 reaches 32.15 compared with 30.74 for Stage~2, and it also moves above Qwen2.5-Omni-7B~\cite{xu2025qwen25omnireport} and Qwen3-Omni-30B-A3B-Instruct~\cite{xu2025qwen3omnireport} under this retained-query mixture. This reversal is mainly because large retained subsets such as AV-Odyssey receive more weight. Since AV-Odyssey alone contributes 4,555 of the 8,551 retained queries, the query-weighted average should not replace the benchmark-level macro average; instead, we treat it as a complementary view of the retained-query mixture.

\begin{figure}[!htbp]
    \centering
    \includegraphics[width=0.92\linewidth]{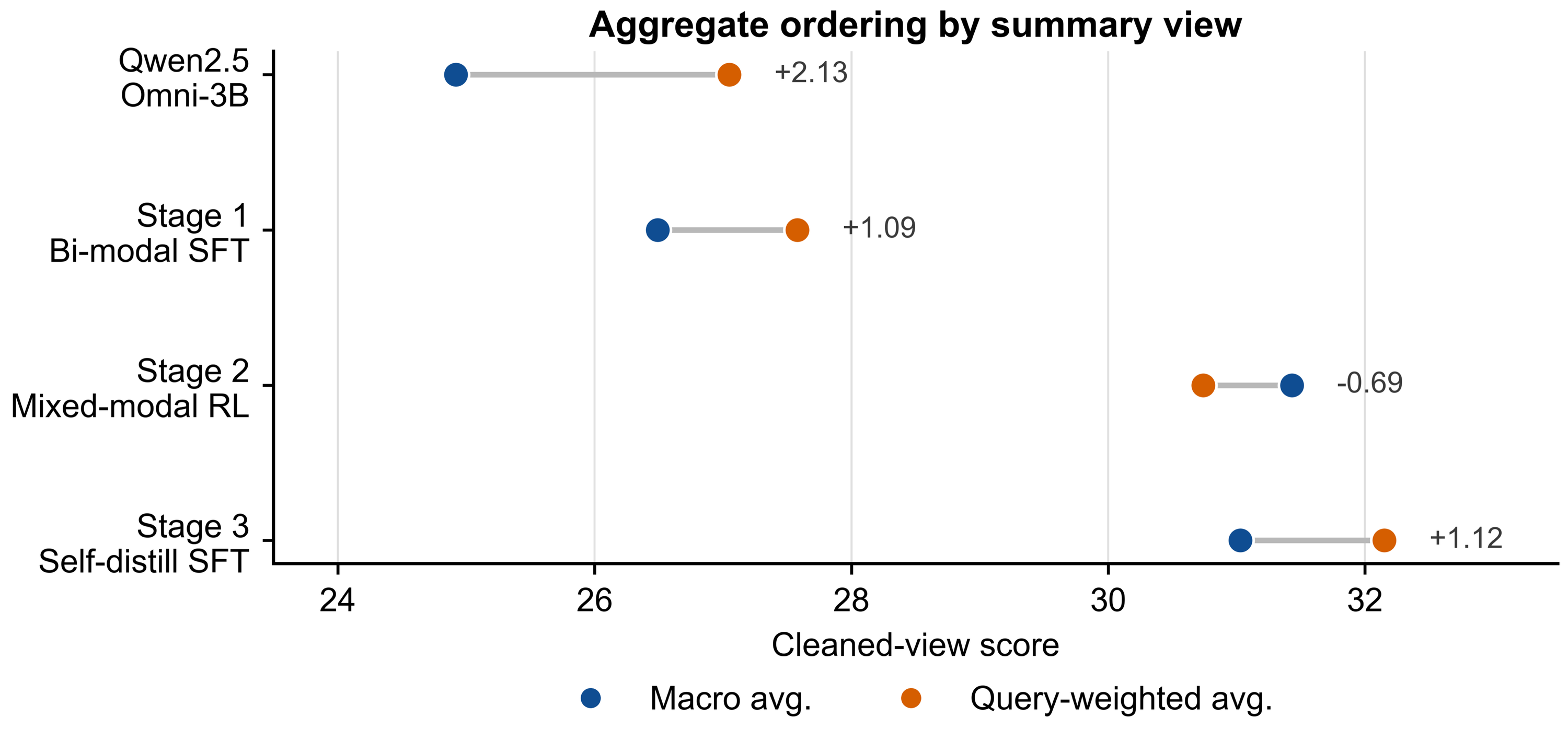}
    \caption{Macro and query-weighted summaries for Qwen2.5-Omni-3B~\cite{xu2025qwen25omnireport} and the three OmniBoost stages. Stage~2 is strongest under the benchmark-level macro average, whereas Stage~3 is strongest under the query-weighted average because large retained subsets, especially AV-Odyssey~\cite{gong2024avodysseybenchmultimodalllms}, receive more weight.}
    \label{fig:omniboost_aggregate_ordering}
\end{figure}

\paragraph{Self-distillation interpretation.} The Stage~3 results indicate that self-distillation is useful but profile-dependent. Because the traces are generated by the same 3B RLVR lineage rather than by a stronger external omni teacher, self-distillation mainly stabilizes and amplifies reasoning patterns already exposed by RLVR. The gains on AV-Odyssey~\cite{gong2024avodysseybenchmultimodalllms}, Daily-Omni~\cite{zhou2026dailyomniaudiovisualreasoningtemporal}, IntentBench~\cite{yang2025humanomniv2understandingomnimodalreasoning}, OmniVideoBench~\cite{li2026omnivideobenchaudiovisualunderstandingevaluation}, and UNO-Bench~\cite{chen2025unobenchunifiedbenchmarkexploring}, together with the query-weighted improvement, show that synthetic queries built from paired audio and video captions provide effective supervision. However, the macro ordering and benchmark-level variation show that filtering and data-ratio choices still matter, so we use the following ablation to isolate the contribution of the filtered synthetic data.

\subsection{Data-Centric Self-Distillation Filtering Ablation}

\paragraph{Fixed ablation setup.} In the fixed comparison below, we use a common SFT setup to compare the value of the self-distillation datasets retained after different filtering passes. Each run starts from the same \textbf{Qwen2.5-Omni-3B} baseline and is fine-tuned with one pass-specific synthetic dataset only. Sequences are packed to 64K tokens; each run trains for 60 steps with a learning rate of $1\times10^{-5}$. This design is intentionally different from the main Stage~3 OmniBoost result, which starts from the RLVR checkpoint and additionally adjusts the data ratio. Table~\ref{tab:self_distill_stage_results} should therefore be read as a data-centric ablation, not as the training trajectory used to produce the Stage~3 column in Table~\ref{tab:sec4_stage_results}.

\newcommand{\gain}[1]{{\footnotesize\textcolor{green}{(+#1)}}}
\newcommand{\loss}[1]{{\footnotesize\textcolor{red}{(-#1)}}}
\newcommand{\rotcol}[1]{\rotatebox[origin=l]{45}{\scriptsize\textbf{#1}}}
\newcommand{\deltacell}[2]{\makecell[c]{#1\\[-1pt]#2}}
\newcommand{\basecell}[1]{\makecell[c]{#1\\[-1pt]\vphantom{\gain{0.00}}}}

\begin{table}[ht]
    \centering
    \caption{Effect of SFT on synthetic datasets retained after each progressive self-distillation filtering pass~\cite{wu2025sdrt}. F1--F3 are cumulative data-filtering passes, not OmniBoost training stages; each run starts from the same Qwen2.5-Omni-3B baseline~\cite{xu2025qwen25omnireport} and uses the fixed setup described in the text. Colored deltas are relative to that baseline; benchmark sources are cited in Section~3.}
    \label{tab:self_distill_stage_results}
    \resizebox{\textwidth}{!}{
    \begin{tabular}{l|ccccccccccc}
        \toprule[1.2pt]
        \textbf{Variant}
        & \rotcol{AV-Odyssey}
        & \rotcol{CG-AV Counting}
        & \rotcol{Daily-Omni}
        & \rotcol{IntentBench}
        & \rotcol{OmniBench}
        & \rotcol{OmniVideoBench}
        & \rotcol{UNO-Bench}
        & \rotcol{Video-Holmes}
        & \rotcol{WorldSense}
        & \rotcol{Macro Avg.}
        & \rotcol{Query-Weighted Avg.} \\
        \midrule
        \textbf{Qwen2.5-Omni-3B} & \basecell{29.00} & \basecell{12.73} & \basecell{27.53} & \basecell{29.57} & \basecell{27.14} & \basecell{27.67} & \basecell{21.41} & \basecell{24.36} & \basecell{24.91} & \basecell{24.92} & \basecell{27.05} \\
        \midrule
        \textbf{SFT on F1-retained Data} & \deltacell{28.47}{\loss{0.53}} & \deltacell{15.16}{\gain{2.43}} & \deltacell{30.38}{\gain{2.85}} & \deltacell{31.06}{\gain{1.49}} & \deltacell{29.74}{\gain{2.60}} & \deltacell{23.90}{\loss{3.77}} & \deltacell{25.44}{\gain{4.03}} & \deltacell{34.46}{\gain{10.10}} & \deltacell{23.09}{\loss{1.82}} & \deltacell{26.86}{\gain{1.94}} & \deltacell{28.02}{\gain{0.97}} \\
        \textbf{SFT on F2-retained Data} & \deltacell{28.96}{\loss{0.04}} & \deltacell{14.36}{\gain{1.63}} & \deltacell{34.60}{\gain{7.07}} & \deltacell{28.64}{\loss{0.93}} & \deltacell{29.50}{\gain{2.36}} & \deltacell{25.79}{\loss{1.88}} & \deltacell{28.95}{\gain{7.54}} & \deltacell{36.38}{\gain{12.02}} & \deltacell{25.60}{\gain{0.69}} & \deltacell{28.09}{\gain{3.17}} & \deltacell{28.78}{\gain{1.74}} \\
        \textbf{SFT on F3-retained Data} & \deltacell{30.03}{\gain{1.03}} & \deltacell{15.69}{\gain{2.96}} & \deltacell{32.07}{\gain{4.54}} & \deltacell{30.75}{\gain{1.18}} & \deltacell{28.78}{\gain{1.64}} & \deltacell{22.33}{\loss{5.34}} & \deltacell{25.88}{\gain{4.47}} & \deltacell{31.98}{\gain{7.62}} & \deltacell{26.29}{\gain{1.38}} & \deltacell{27.09}{\gain{2.17}} & \deltacell{28.87}{\gain{1.83}} \\
        \bottomrule[1.2pt]
    \end{tabular}}
\end{table}

\paragraph{Filtering-pass comparison.} Table~\ref{tab:self_distill_stage_results} shows that directly applying SFT with any pass-retained self-distillation dataset improves Qwen2.5-Omni-3B~\cite{xu2025qwen25omnireport} under both aggregate views, confirming that the synthetic supervision is useful even without the full staged recipe. F2-retained data gives the strongest macro average, while F3-retained data is only slightly stronger under the query-weighted average. The gains remain benchmark-dependent: Video-Holmes~\cite{cheng2025videoholmesmllmthinklike}, Daily-Omni~\cite{zhou2026dailyomniaudiovisualreasoningtemporal}, and UNO-Bench~\cite{chen2025unobenchunifiedbenchmarkexploring} improve most clearly, whereas OmniVideoBench~\cite{li2026omnivideobenchaudiovisualunderstandingevaluation} declines under all three fixed ablation datasets. This table is therefore a data-centric ablation of filtered synthetic supervision, not the final Stage~3 training trajectory.

\section{Conclusion}

This work shows that visually answerable queries can make omni-modal benchmarks overstate omni-modal understanding, and introduces \textbf{OmniClean} as a visually debiased evaluation view built by query-level visual-only probing over nine existing benchmarks. On this cleaned view, \textbf{OmniBoost} shows that balanced mixed bi-modal supervised fine-tuning~\cite{10.5555/3600270.3602281,wang2022selfinstruct,liu2023visual} is a useful control but is not sufficient for consistent omni-modal gains, while mixed-modality RLVR~\cite{shao2024deepseekmath,deepseekai2025,yu2025dapoopensource} provides the clearest benchmark-level macro improvement. Self-distillation~\cite{hinton2015distilling,wu2025sdrt} remains useful but profile-dependent: Stage~3 leads under query-weighted aggregation and the fixed ablations show that synthetic queries built from paired audio and video captions can directly improve the 3B base model, but it does not uniformly dominate RLVR across benchmark families. These findings support a practical conclusion: progress in omni-modal language models is easier to interpret when evaluation first controls visual leakage, and small models can benefit substantially from staged post-training with explicitly constructed omni-modal supervision, with the present evidence scoped to one Qwen2.5-Omni-3B lineage~\cite{xu2025qwen25omnireport} and our visual-only leakage protocol.

\section*{Author List}

\noindent\textbf{StepFun-Audio Team}

\vspace{0.5em}

\noindent
Che Liu$^{1,2}$, Lichao Ma$^{1,3}$, Xiangyu Tony Zhang$^{1,5}$,
Yuxin Zhang$^{1,4}$, Haoyang Zhang$^{1,3}$, Xuerui Yang$^{1}$, and Fei Tian$^{1,*}$.

\vspace{0.6em}

\noindent{\normalsize
$^{1}$StepFun; 
$^{2}$Imperial College London; 
$^{3}$Peking University; 
$^{4}$Shanghai Jiao Tong University; 
$^{5}$The University of New South Wales.
}

\vspace{0.5em}

\noindent{\small $^{*}$Corresponding authors: \texttt{tianfei@stepfun.com}}

\vspace{0.5em}
{\footnotesize
\setlength{\bibsep}{0.2\baselineskip}
\bibliography{paper}
}

\appendix

\makeatletter
\setlength{\@fptop}{0pt}
\makeatother

\section{Detailed Synthetic Query Graphic Description}
\label{appendix:synthetic_query_graphic_description}

\begin{figure}[!htbp]
    \centering
    \includegraphics[width=\textwidth]{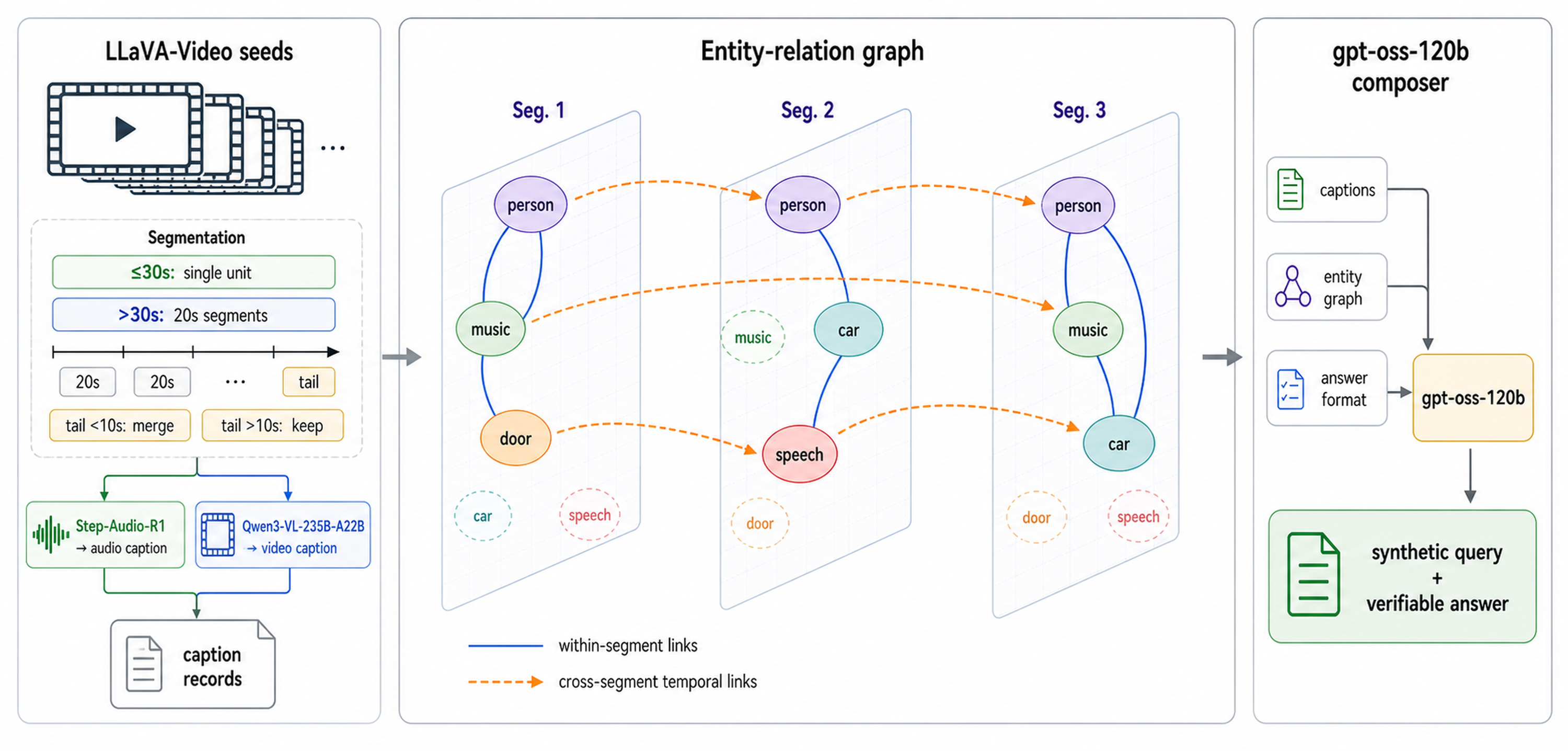}
    \caption{Detailed graphic description of the Synthetic Query construction process. The left panel expands the seed-video segmentation rule and caption-record construction, the middle panel illustrates the entity-relation scaffold with within-segment and cross-segment temporal links, and the right panel shows how captions, the entity graph, and answer-format constraints are provided to gpt-oss-120b~\cite{openai2025gptoss} to compose a Synthetic Query with a verifiable answer. This appendix graphic describes the same process summarized compactly in Figure~\ref{fig:synthetic_query_construction}.}
    \label{fig:synthetic_query_graphic_description}
\end{figure}

\section{Full Section 3 Regression Plots}
\label{appendix:full_unimodal_regression}

\begin{center}
    \centering
    \begin{minipage}[t]{0.47\textwidth}
        \centering
        \includegraphics[width=0.96\textwidth]{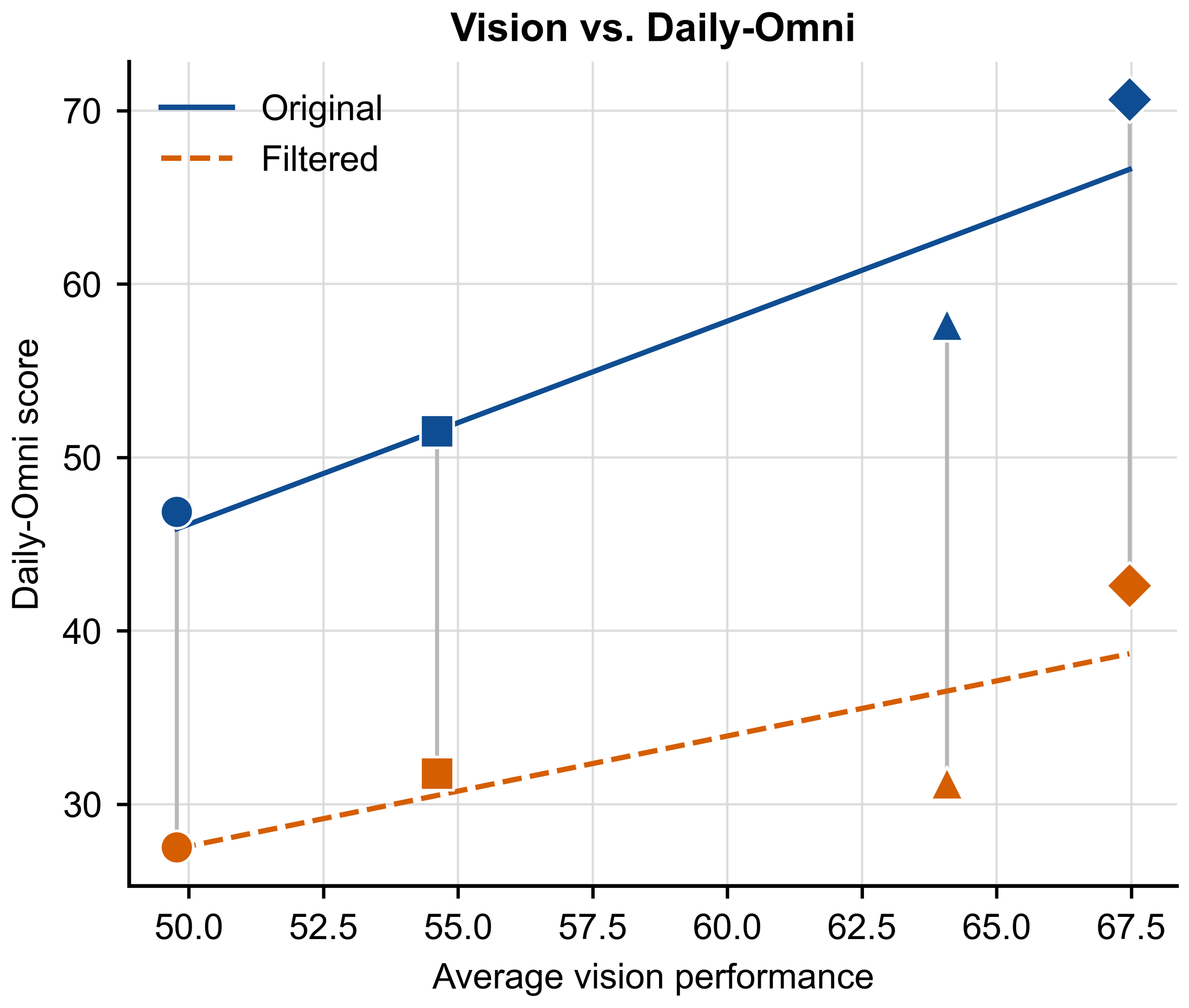}

        {\scriptsize (a) Daily-Omni: Vision}
    \end{minipage}
    \hfill
    \begin{minipage}[t]{0.47\textwidth}
        \centering
        \includegraphics[width=0.96\textwidth]{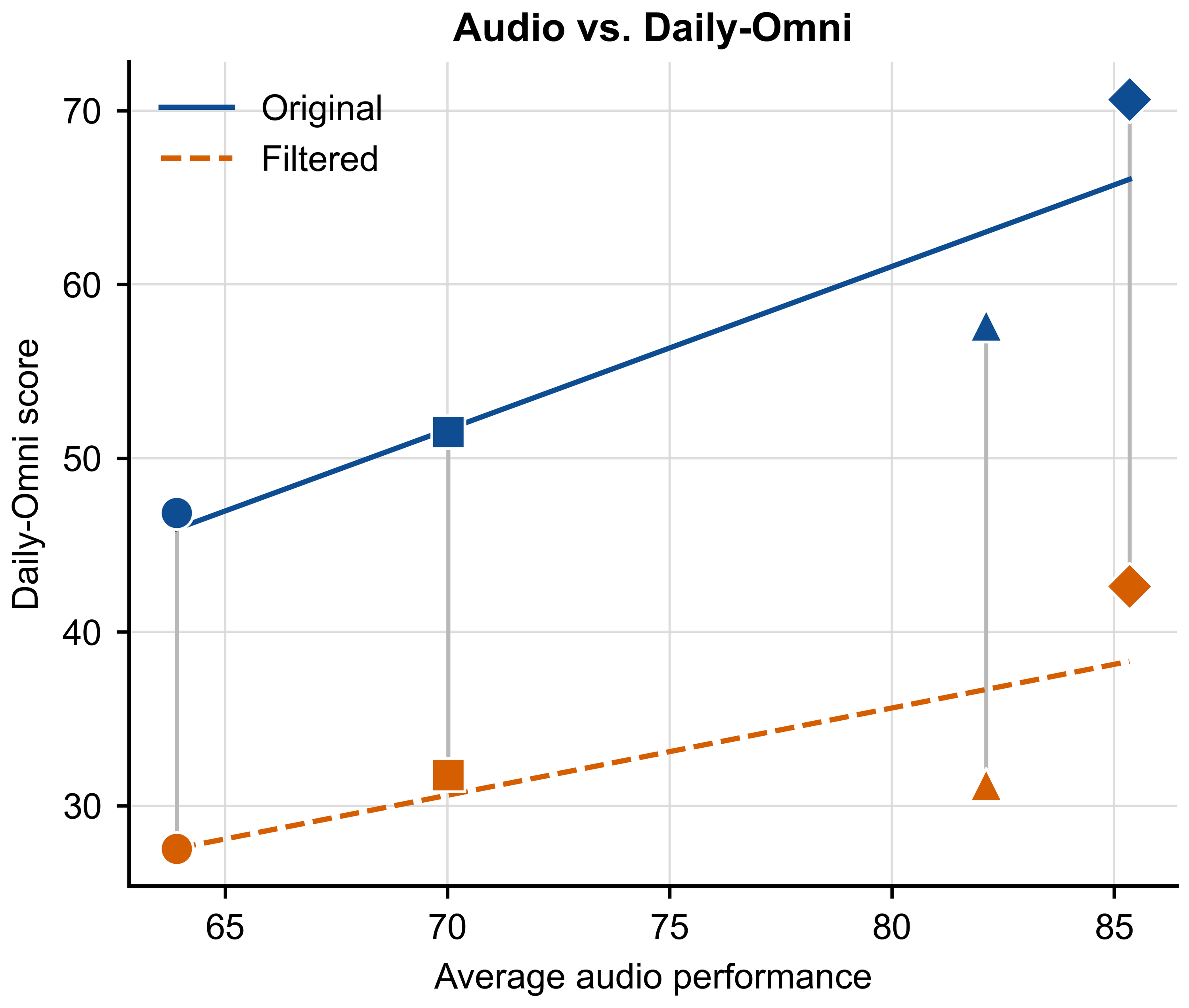}

        {\scriptsize (b) Daily-Omni: Audio}
    \end{minipage}

    \par\vspace{0.1em}

    \begin{minipage}[t]{0.47\textwidth}
        \centering
        \includegraphics[width=0.96\textwidth]{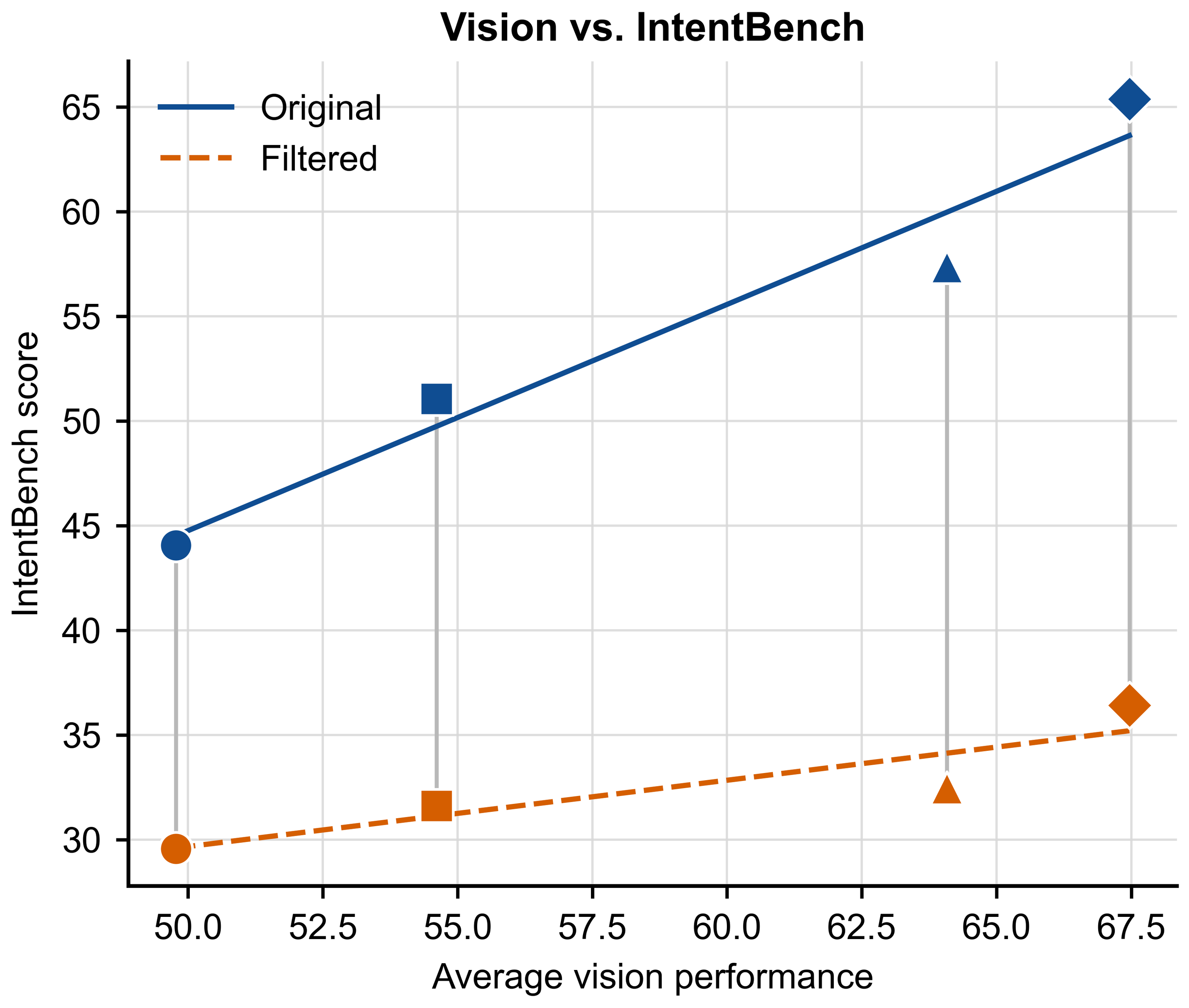}

        {\scriptsize (c) IntentBench: Vision}
    \end{minipage}
    \hfill
    \begin{minipage}[t]{0.47\textwidth}
        \centering
        \includegraphics[width=0.96\textwidth]{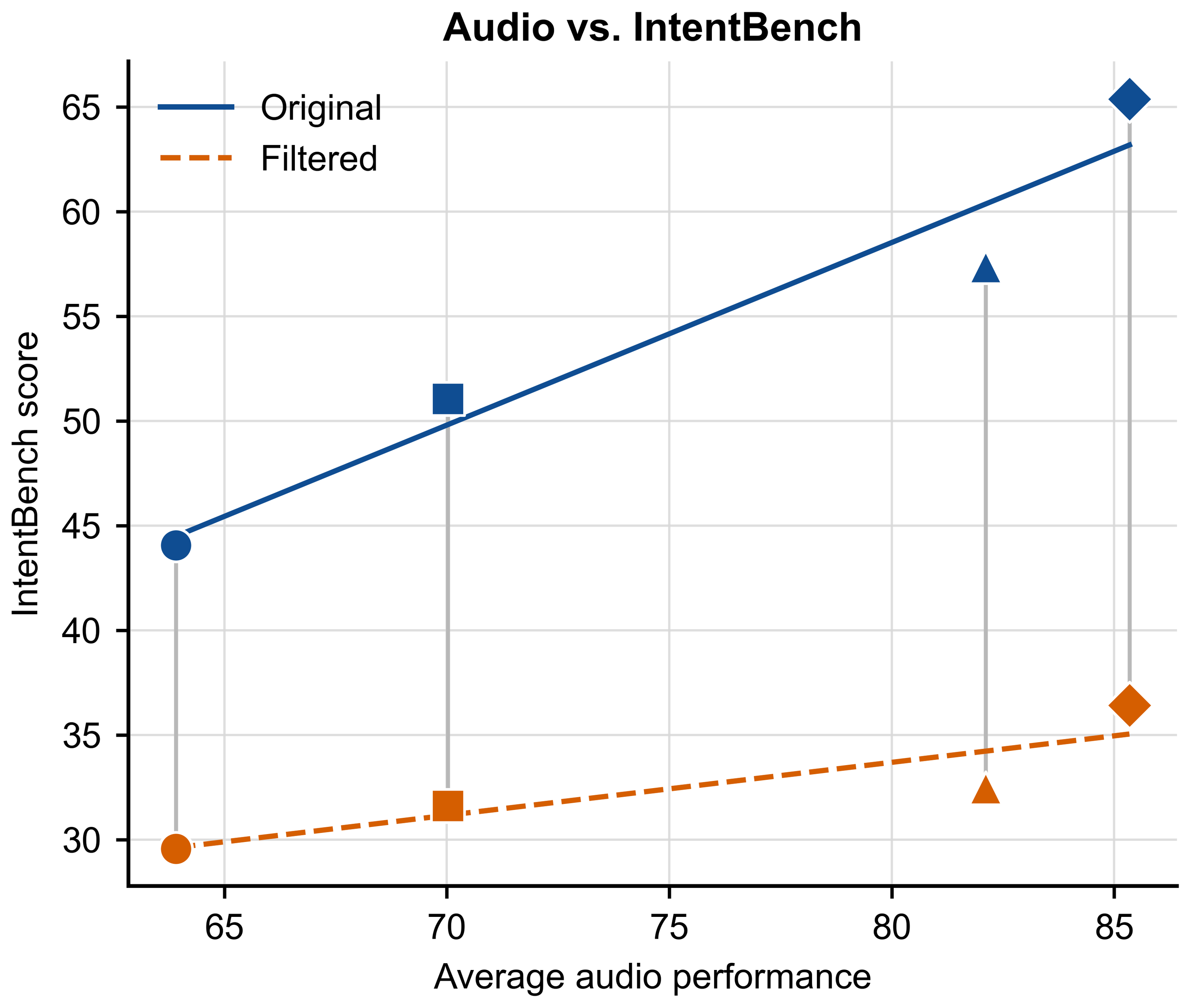}

        {\scriptsize (d) IntentBench: Audio}
    \end{minipage}

    \par\vspace{0.1em}

    \begin{minipage}[t]{0.47\textwidth}
        \centering
        \includegraphics[width=0.96\textwidth]{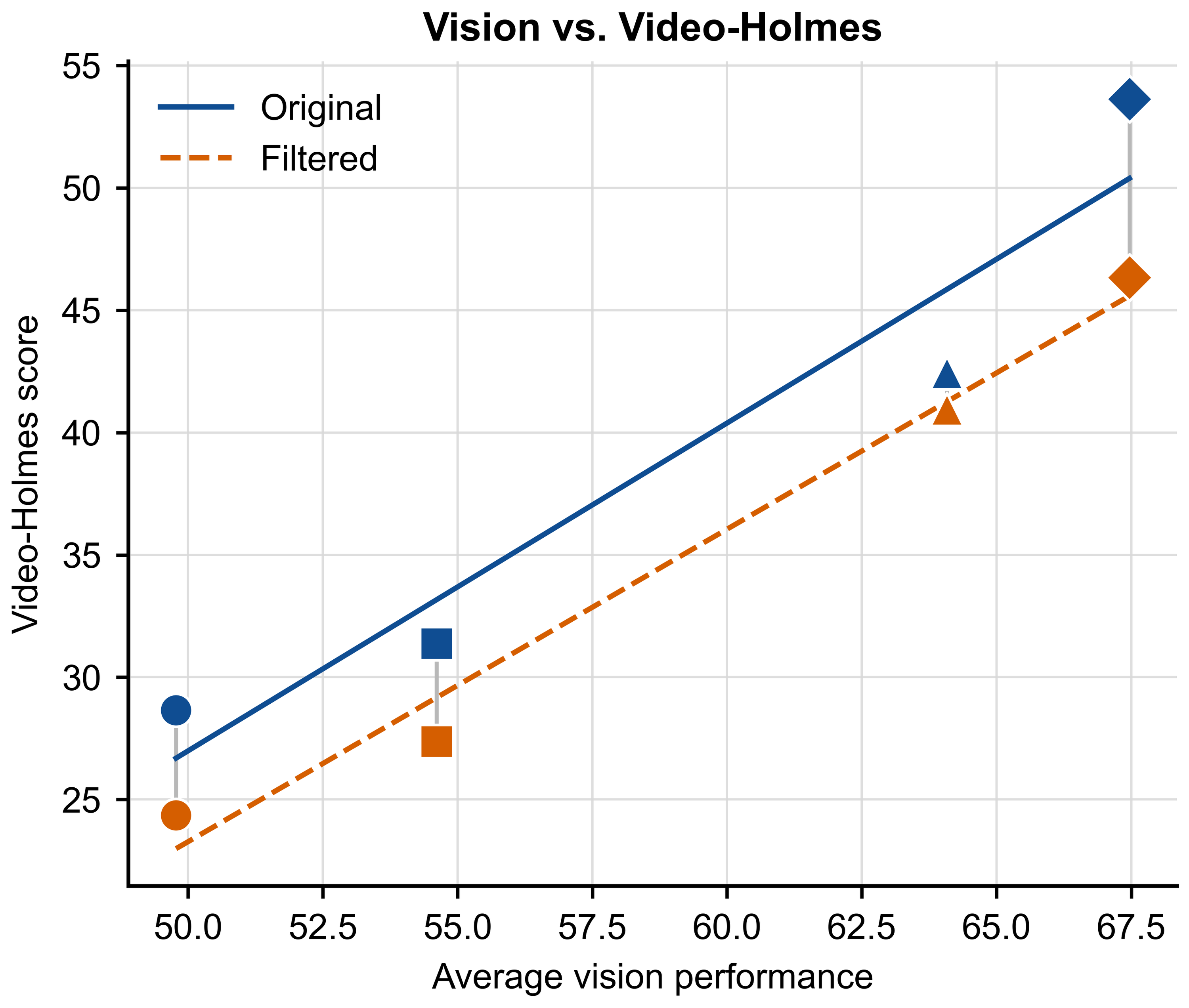}

        {\scriptsize (e) Video-Holmes: Vision}
    \end{minipage}
    \hfill
    \begin{minipage}[t]{0.47\textwidth}
        \centering
        \includegraphics[width=0.96\textwidth]{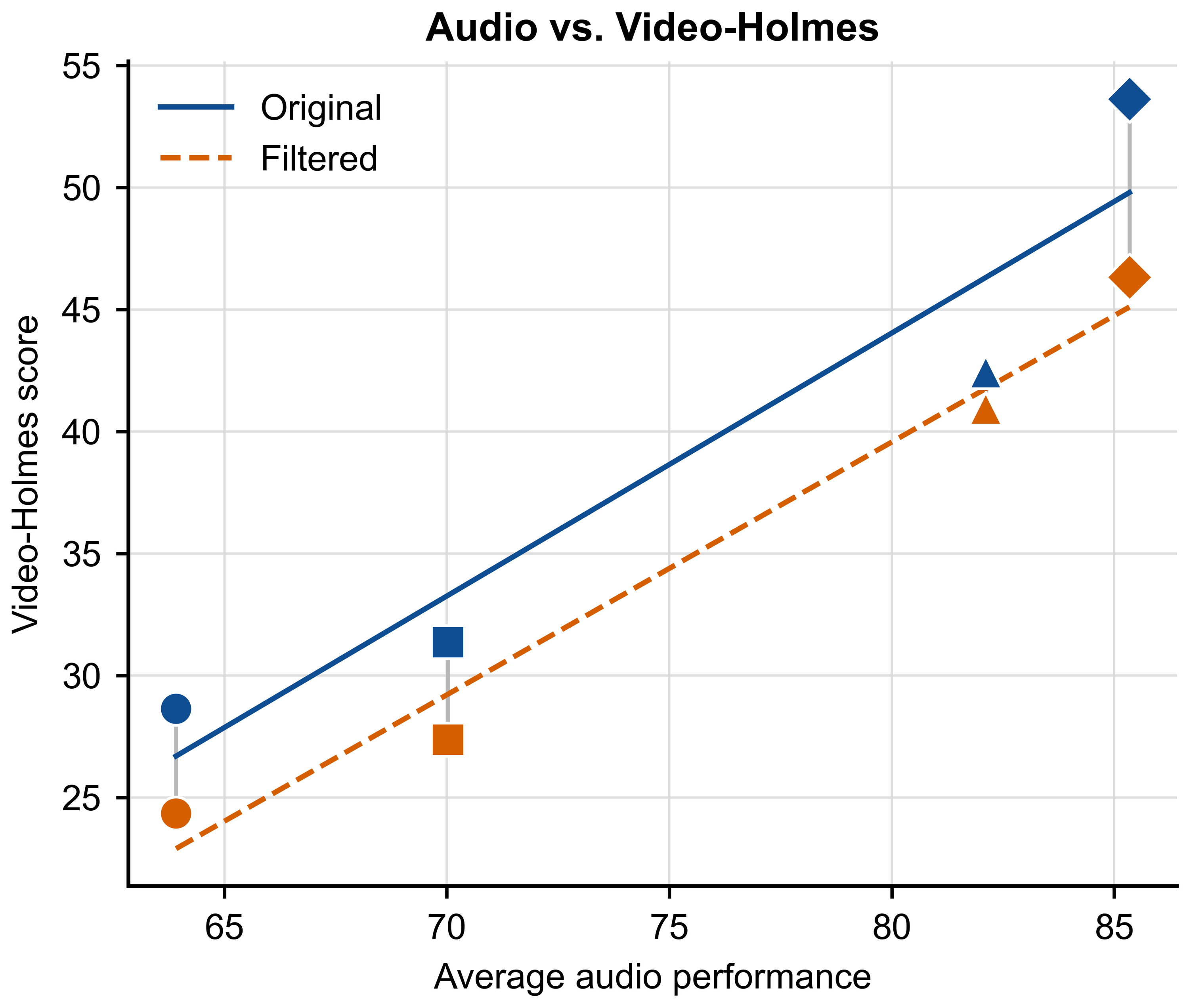}

        {\scriptsize (f) Video-Holmes: Audio}
    \end{minipage}

    \captionsetup{hypcap=false}
    \captionof{figure}{Benchmark-by-benchmark regression panels for Daily-Omni~\cite{zhou2026dailyomniaudiovisualreasoningtemporal}, IntentBench~\cite{yang2025humanomniv2understandingomnimodalreasoning}, and Video-Holmes~\cite{cheng2025videoholmesmllmthinklike}. Each dataset is shown with paired vision-score and audio-score views against the omni score.}
    \label{fig:unimodal_regression_full_a}
\end{center}

\begin{center}
    \centering

    \begin{minipage}[t]{0.47\textwidth}
        \centering
        \includegraphics[width=0.96\textwidth]{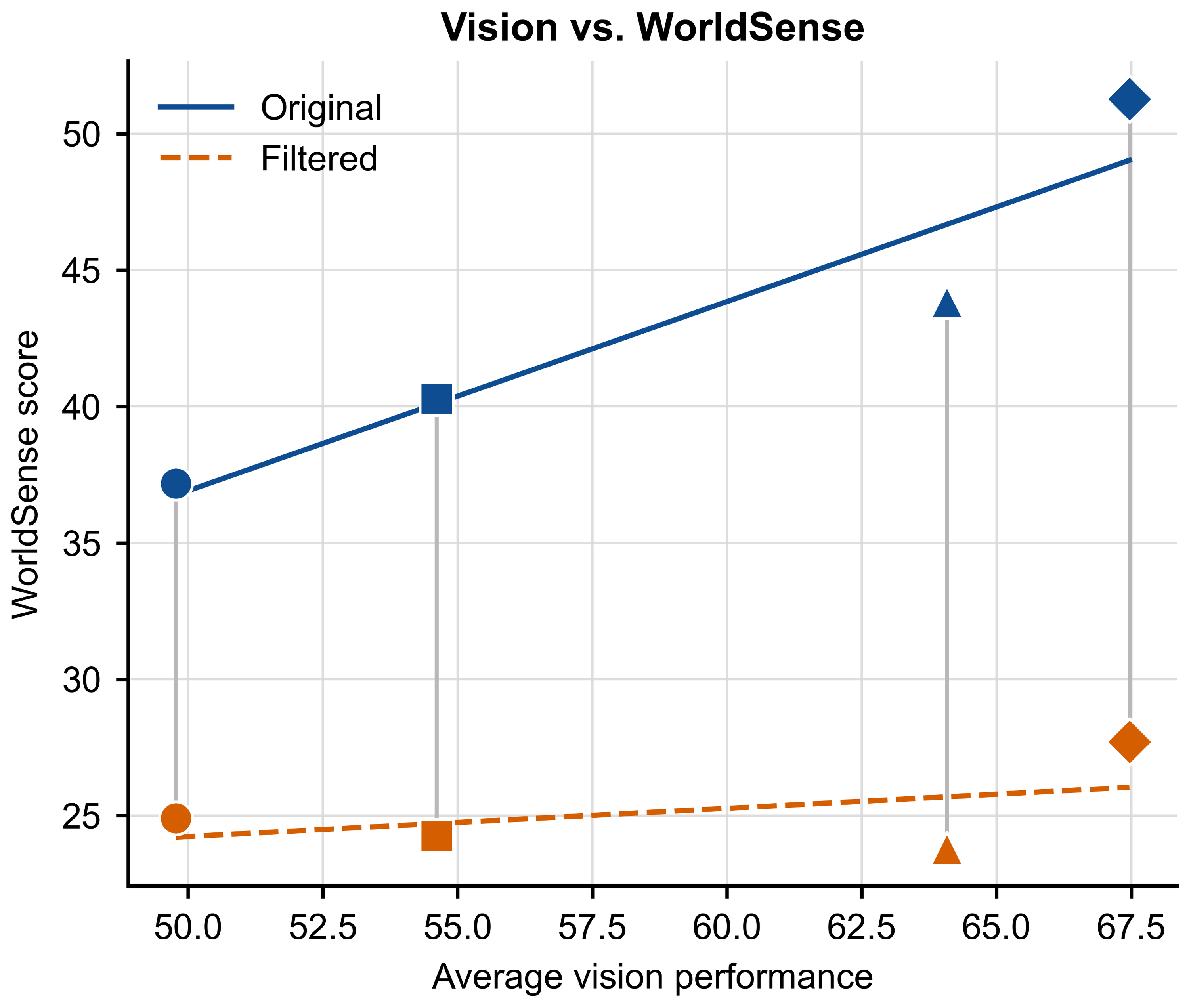}
        
        {\scriptsize (a) WorldSense: Vision}
    \end{minipage}
    \hfill
    \begin{minipage}[t]{0.47\textwidth}
        \centering
        \includegraphics[width=0.96\textwidth]{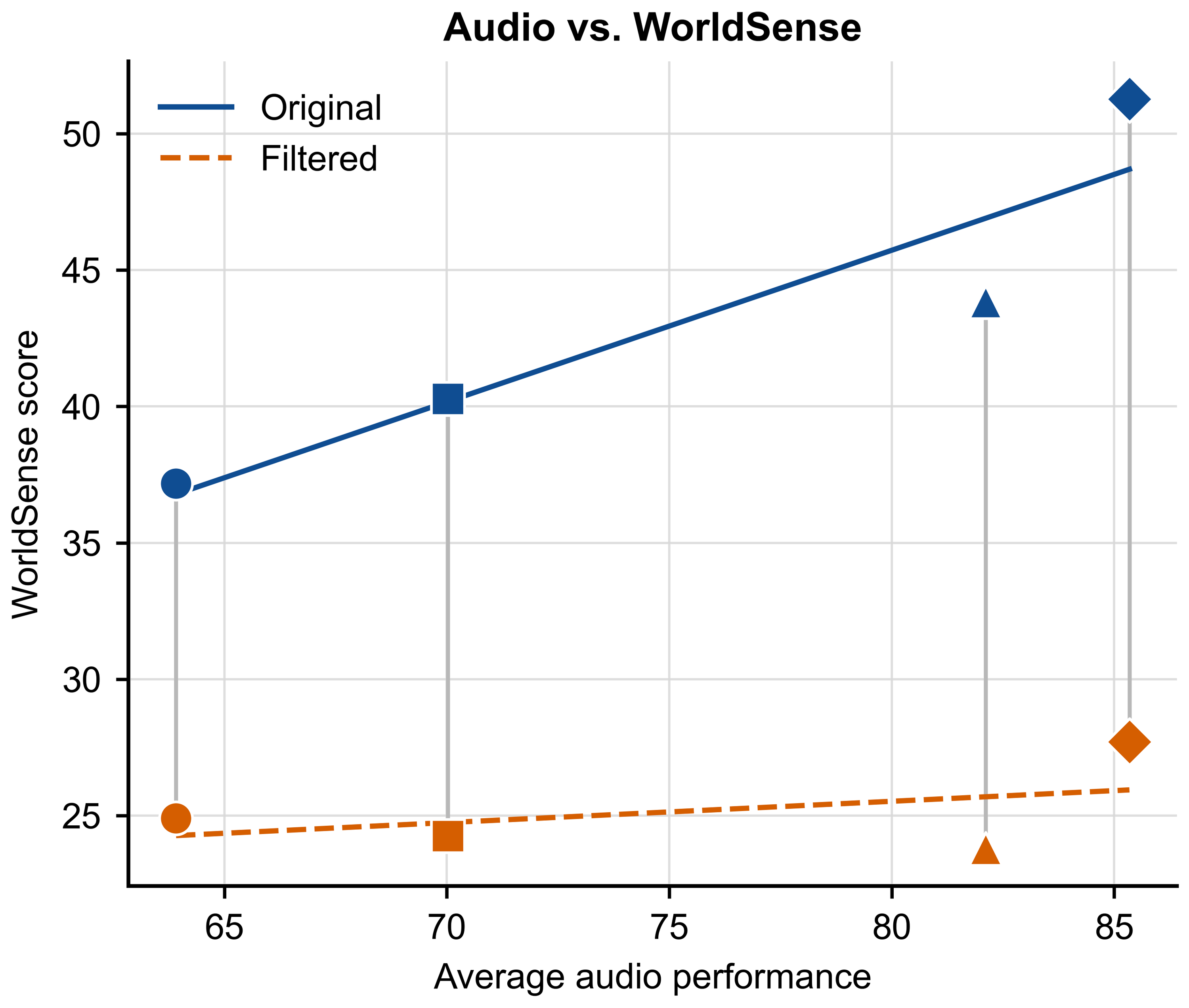}
        
        {\scriptsize (b) WorldSense: Audio}
    \end{minipage}

    \par\vspace{0.1em}

    \begin{minipage}[t]{0.47\textwidth}
        \centering
        \includegraphics[width=0.96\textwidth]{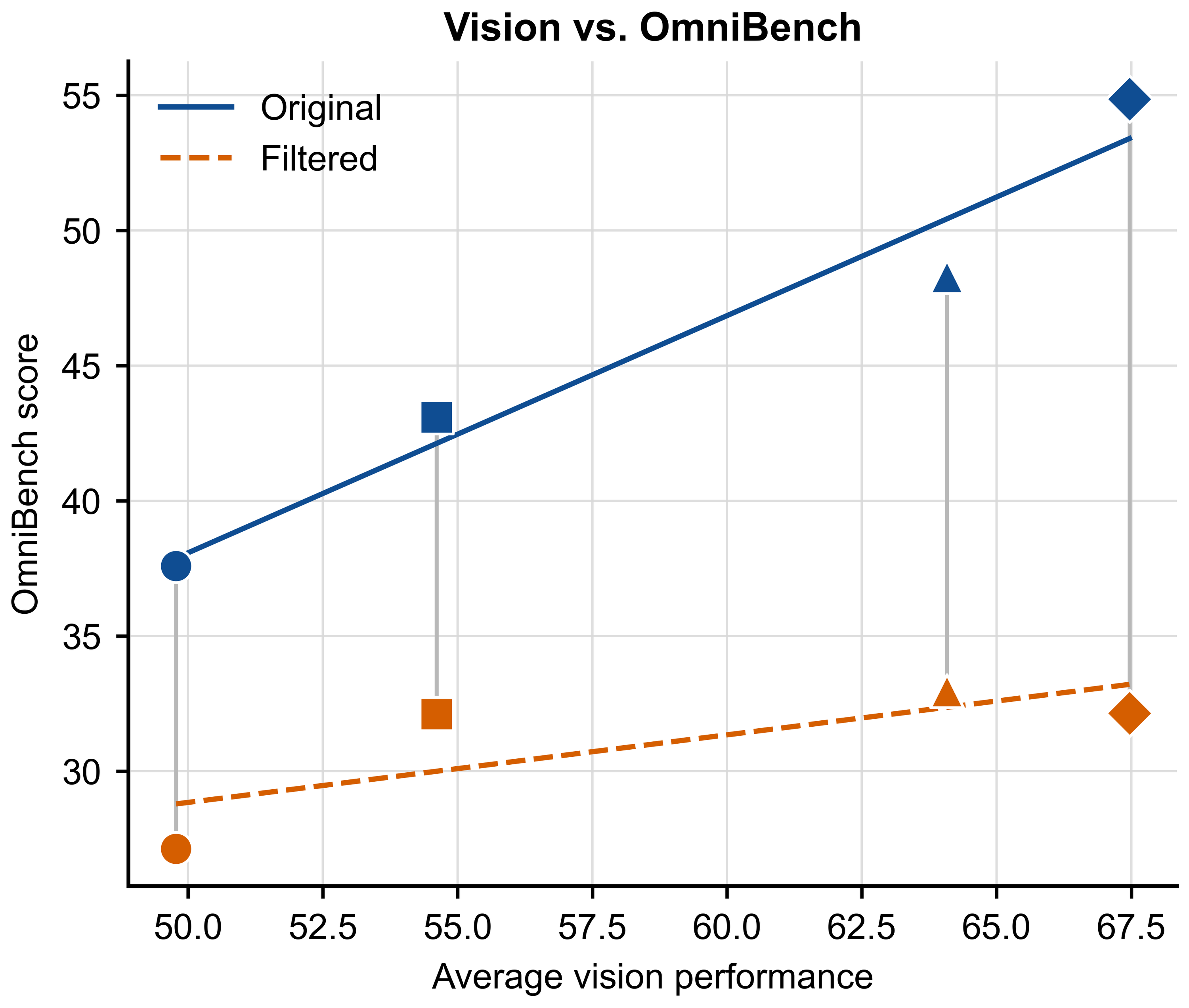}
        
        {\scriptsize (c) OmniBench: Vision}
    \end{minipage}
    \hfill
    \begin{minipage}[t]{0.47\textwidth}
        \centering
        \includegraphics[width=0.96\textwidth]{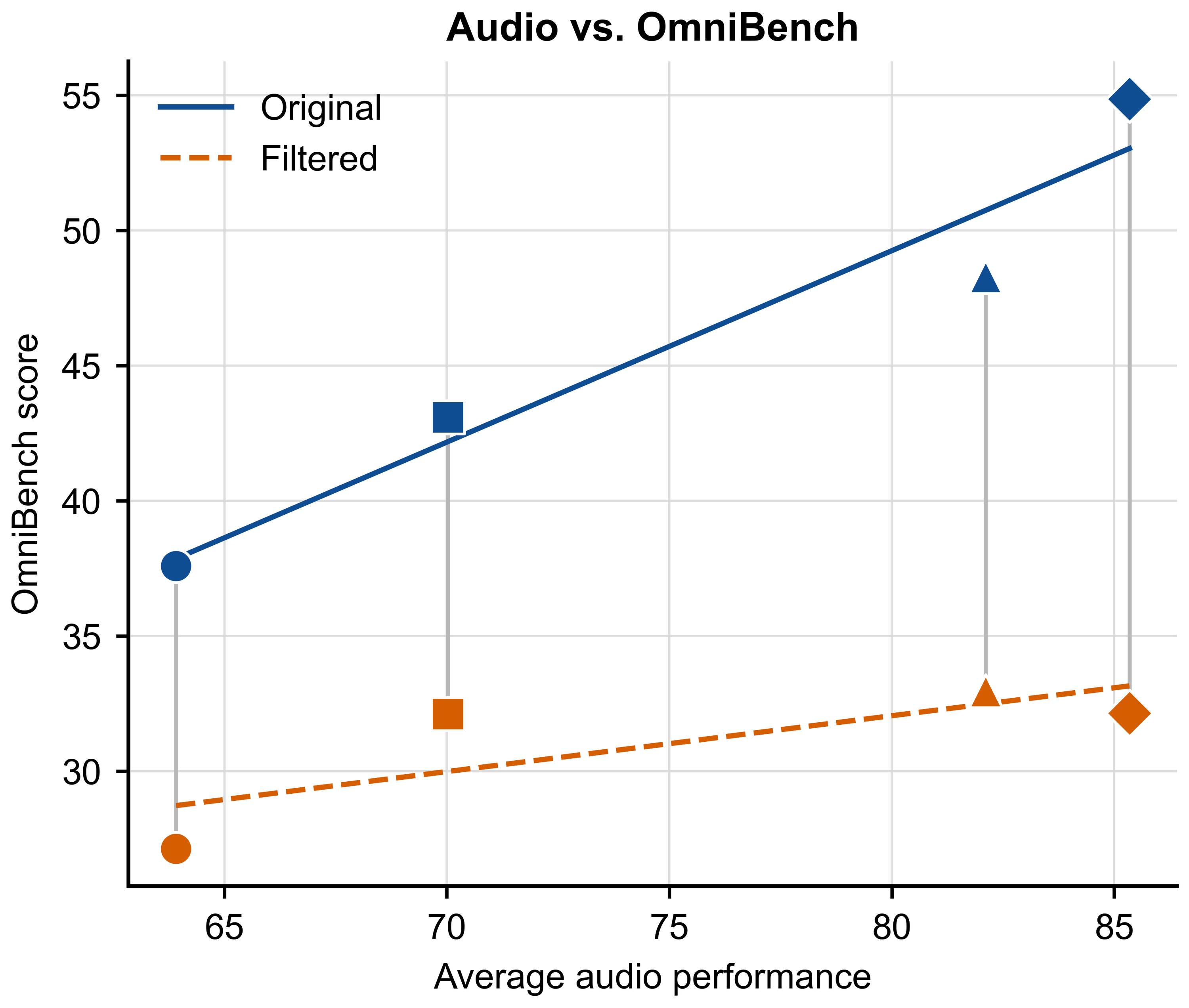}
        
        {\scriptsize (d) OmniBench: Audio}
    \end{minipage}

    \par\vspace{0.1em}

    \begin{minipage}[t]{0.47\textwidth}
        \centering
        \includegraphics[width=0.96\textwidth]{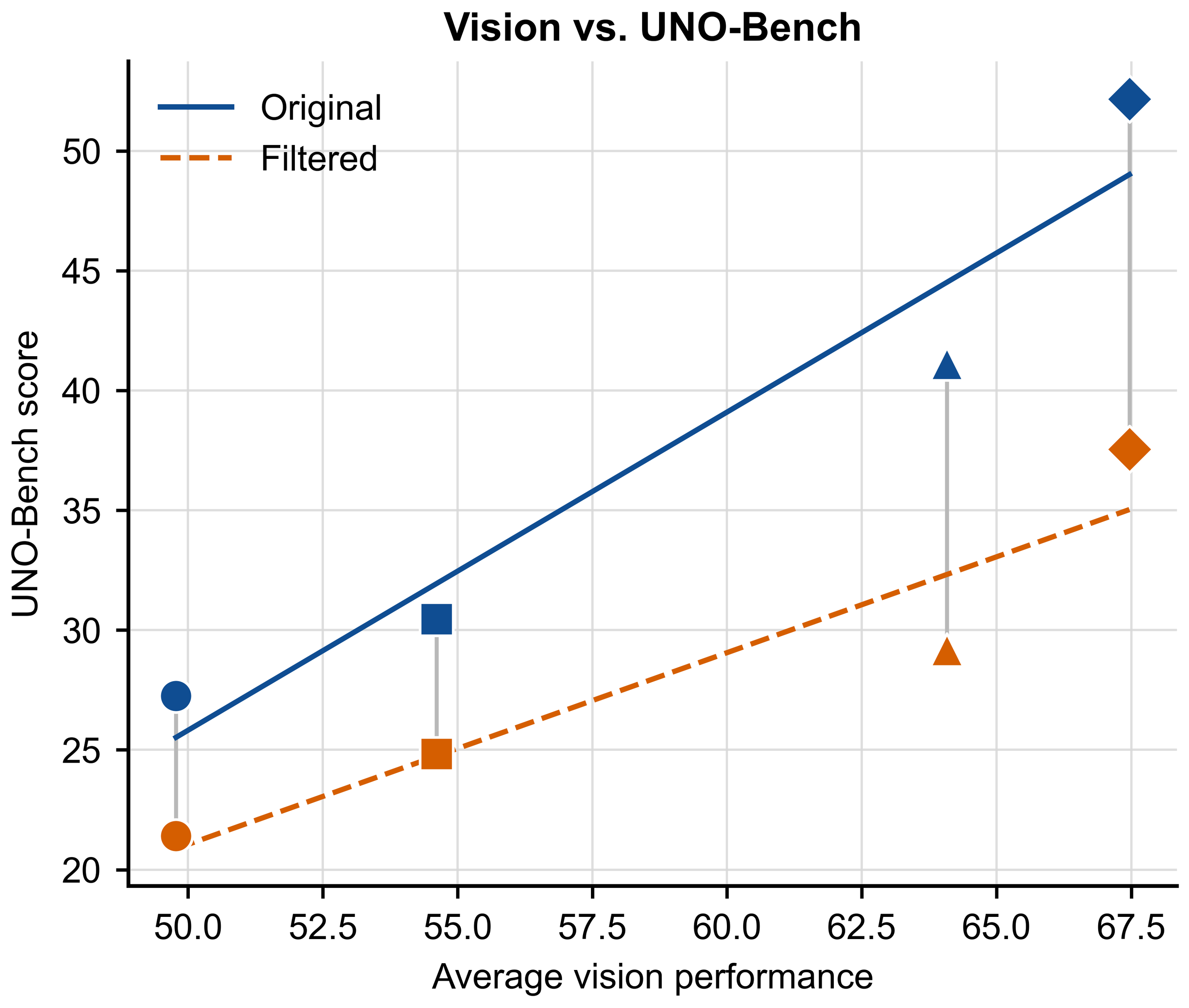}
        
        {\scriptsize (e) UNO-Bench: Vision}
    \end{minipage}
    \hfill
    \begin{minipage}[t]{0.47\textwidth}
        \centering
        \includegraphics[width=0.96\textwidth]{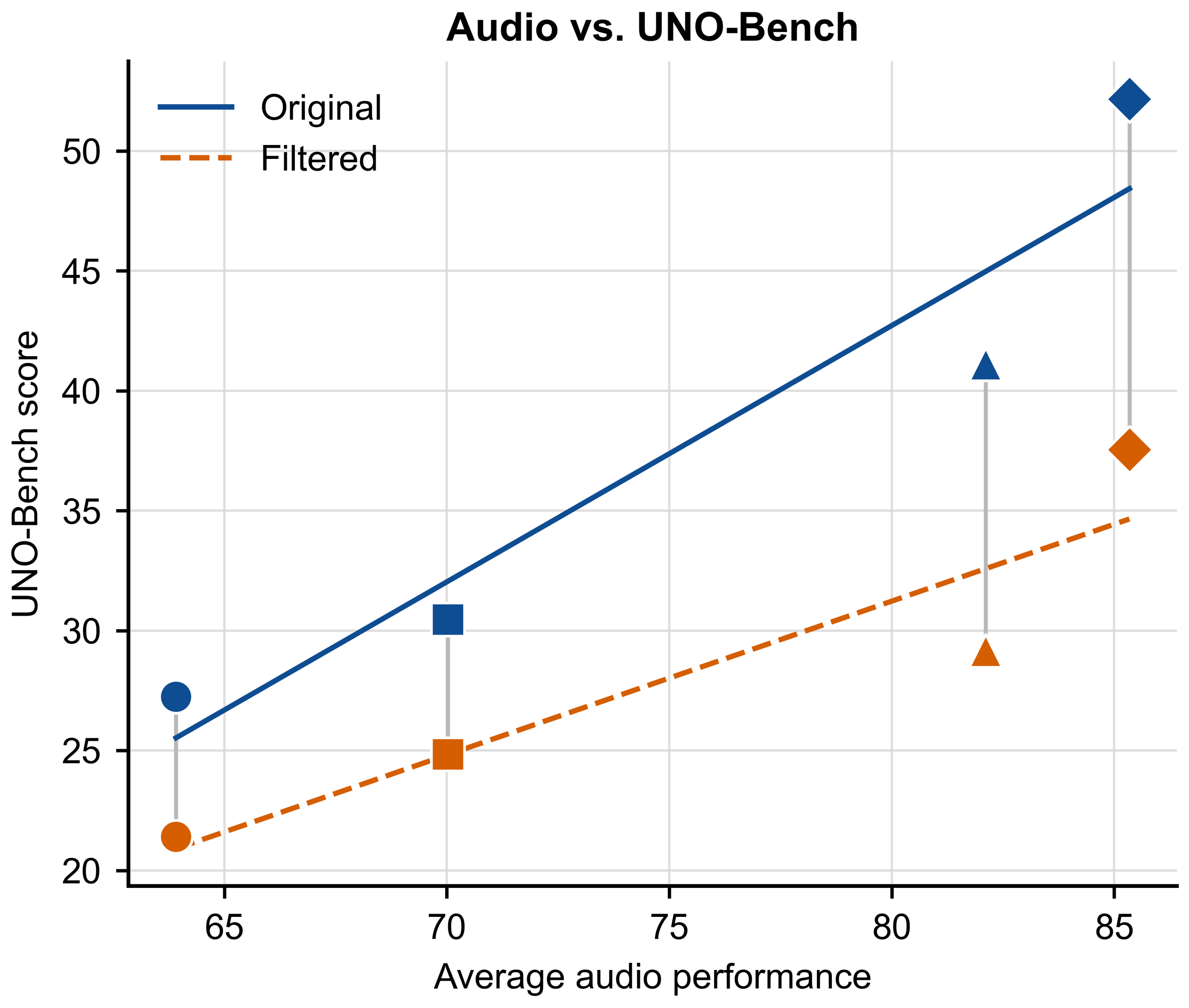}
        
        {\scriptsize (f) UNO-Bench: Audio}
    \end{minipage}

    \captionsetup{hypcap=false}
    \captionof{figure}{Additional benchmark-by-benchmark regression panels for WorldSense~\cite{hong2026worldsenseevaluatingrealworldomnimodal}, OmniBench~\cite{li2025omnibenchfutureuniversalomnilanguage}, and UNO-Bench~\cite{chen2025unobenchunifiedbenchmarkexploring}, continuing the Section~3 gallery.}
    \label{fig:unimodal_regression_full_b}
\end{center}

\begin{center}
    \centering

    \begin{minipage}[t]{0.47\textwidth}
        \centering
        \includegraphics[width=0.96\textwidth]{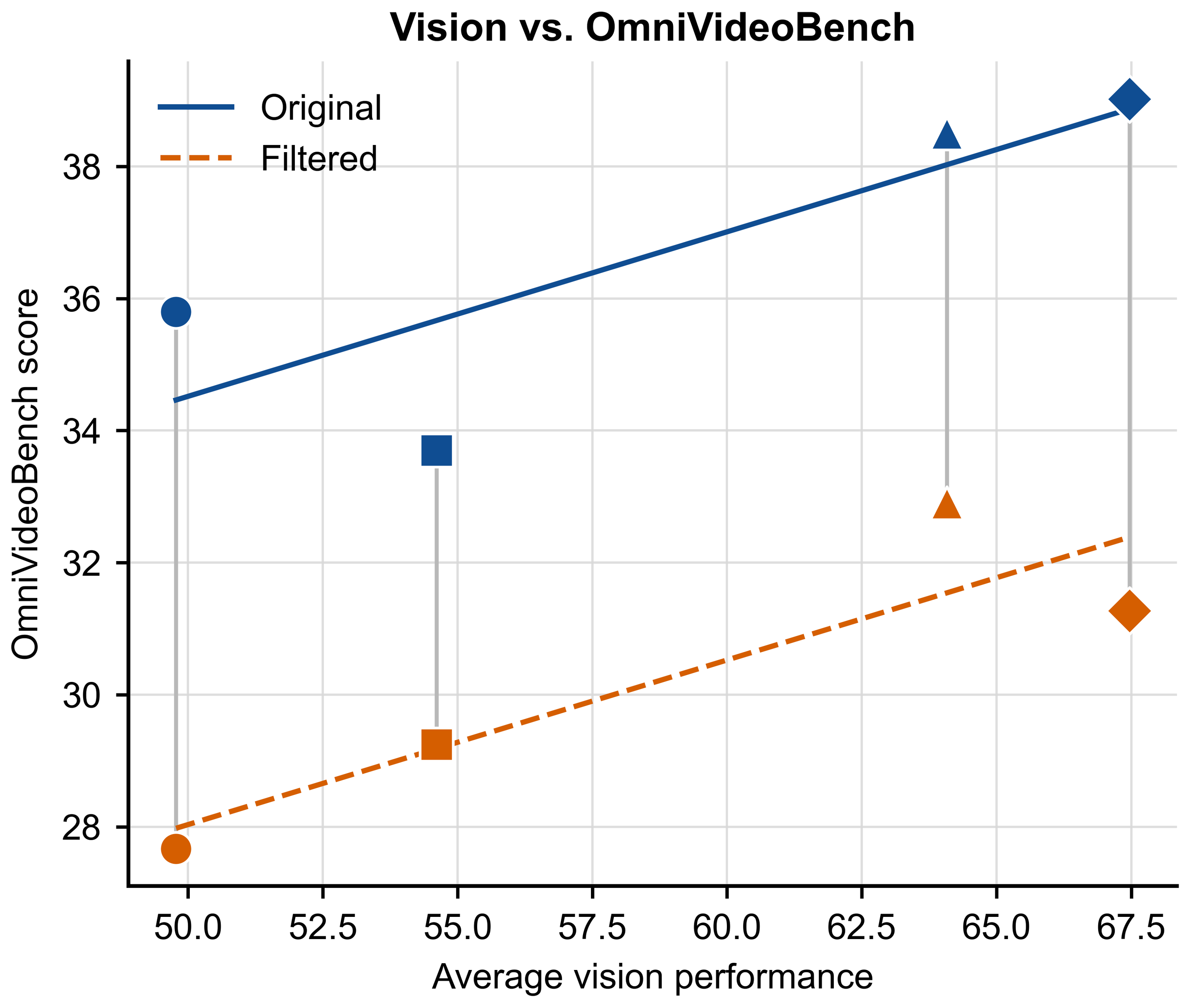}
        
        {\scriptsize (a) OmniVideoBench: Vision}
    \end{minipage}
    \hfill
    \begin{minipage}[t]{0.47\textwidth}
        \centering
        \includegraphics[width=0.96\textwidth]{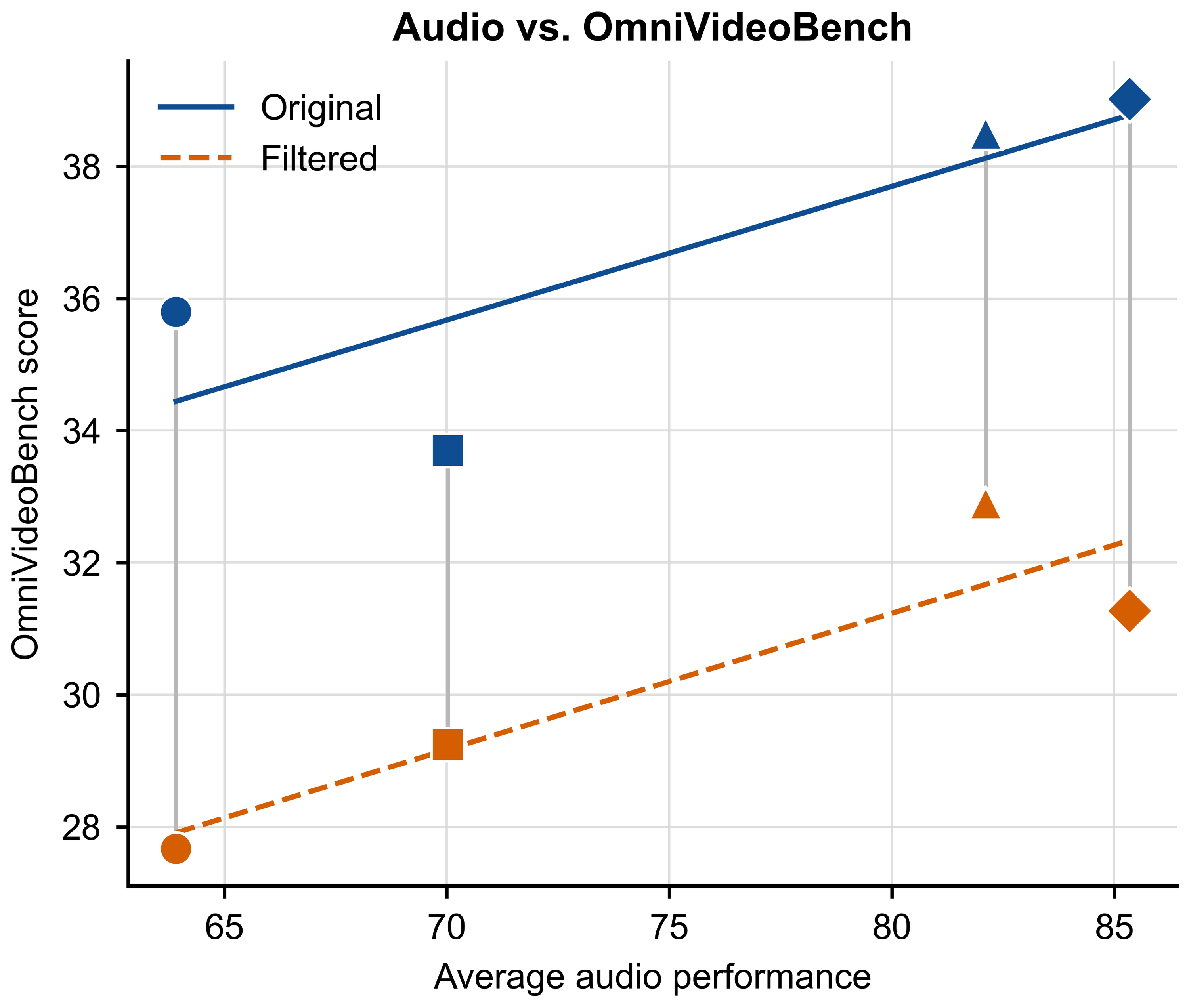}
        
        {\scriptsize (b) OmniVideoBench: Audio}
    \end{minipage}

    \captionsetup{hypcap=false}
    \captionof{figure}{Final regression panels for OmniVideoBench~\cite{li2026omnivideobenchaudiovisualunderstandingevaluation}. AV-Odyssey~\cite{gong2024avodysseybenchmultimodalllms} and CG-AV-Counting~\cite{lu2025avreasonerimprovingbenchmarkingcluegrounded} are omitted because they do not have reported filtered-score views under our protocol.}
    \label{fig:unimodal_regression_full}
\end{center}

\section{Cleaned-View Stage Delta Visualization}
\label{appendix:stage_delta_heatmap}

\begin{center}
    \centering
    \includegraphics[width=0.92\linewidth]{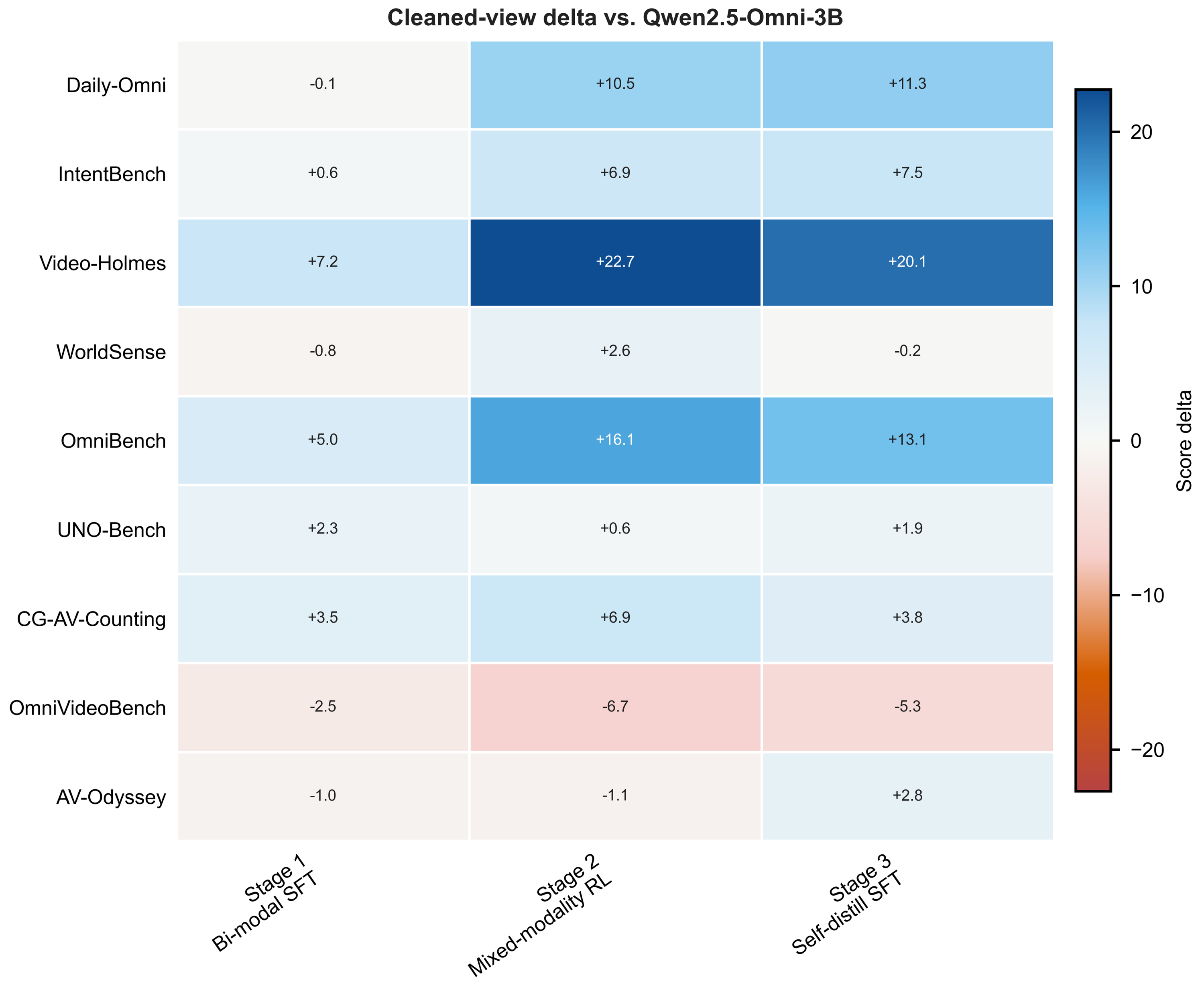}
    \captionof{figure}{Benchmark-level score deltas on the cleaned evaluation view relative to Qwen2.5-Omni-3B~\cite{xu2025qwen25omnireport}. This visual companion to Table~\ref{tab:sec4_stage_results} highlights the redistribution pattern: Stage~2 produces the largest gains on Video-Holmes~\cite{cheng2025videoholmesmllmthinklike}, OmniBench~\cite{li2025omnibenchfutureuniversalomnilanguage}, and Daily-Omni~\cite{zhou2026dailyomniaudiovisualreasoningtemporal}, while Stage~3 shifts strength toward AV-Odyssey~\cite{gong2024avodysseybenchmultimodalllms}, Daily-Omni, IntentBench~\cite{yang2025humanomniv2understandingomnimodalreasoning}, OmniVideoBench~\cite{li2026omnivideobenchaudiovisualunderstandingevaluation}, and UNO-Bench~\cite{chen2025unobenchunifiedbenchmarkexploring}.}
    \label{fig:omniboost_stage_delta_heatmap}
\end{center}

\section{Original-View Results Across the Three OmniBoost Stages}
\label{appendix:original_view_stage_results}

For completeness, we also report original-view scores before query-level cleaning for the same three OmniBoost stages discussed in Section~4. These results are supplementary rather than primary: because the original evaluation view still contains visually answerable queries, it remains affected by visual leakage and is therefore not the main basis for our conclusions. Their purpose is narrower. They help check whether the stage ordering observed on the cleaned evaluation view still appears before cleaning, or whether that ordering is entirely an artifact of the cleaned construction. To make this appendix table self-contained, we also include the original-view scores of the four open-source omni models from Table~\ref{tab:sketch} as reference points.

Table~\ref{tab:appendix_original_stage_results} shows that the same broad pattern remains visible. \textbf{Stage 2: Mixed-Modality RLVR} has the strongest macro and original-weighted averages across the three OmniBoost stages, while \textbf{Stage 1: Mixed Bi-modal SFT} is clearly weaker. \textbf{Stage 3: Self-Distillation SFT} remains useful, but it does not uniformly surpass the RLVR stage: it improves only a subset of datasets, most visibly AV-Odyssey~\cite{gong2024avodysseybenchmultimodalllms} and OmniVideoBench~\cite{li2026omnivideobenchaudiovisualunderstandingevaluation}, while the RLVR stage stays stronger on most other benchmarks. The added reference columns also show that these original-view stage results can look competitive against larger open-source omni models, which is precisely why we do not treat the original view as the main basis for interpretation: once visual leakage remains present, strong scores can still reflect shortcut-sensitive benchmark structure. This appendix result therefore echoes the main text. Even on the original view, RLVR is the first broad-gain stage of OmniBoost, whereas later self-distillation SFT mainly redistributes strengths across benchmarks rather than becoming a uniformly dominant endpoint.

\begin{center}
\centering
\captionof{table}{Original-view results before query-level cleaning. We include both the three OmniBoost stages and the four open-source omni reference models from Table~\ref{tab:sketch}; the reference columns use Qwen2.5-Omni~\cite{xu2025qwen25omnireport} and Qwen3-Omni~\cite{xu2025qwen3omnireport}. Benchmark sources are cited in Section~3. Macro averages weight benchmarks equally; original-weighted averages summarize the original-view mixture represented by this table. Averages are computed from the original-query counts and unrounded scores. These scores are reported only as supplementary context because the original evaluation view is still affected by visual leakage.}
\label{tab:appendix_original_stage_results}
\small
\resizebox{\textwidth}{!}{
\begin{tabular}{l|cccc|ccc}
\toprule[1.2pt]
\textbf{Benchmark}
& \shortstack[c]{\textbf{Qwen2.5-Omni} \\ \textbf{3B}}
& \shortstack[c]{\textbf{Qwen2.5-Omni} \\ \textbf{7B}}
& \shortstack[c]{\textbf{Qwen3-Omni} \\ \textbf{30B-A3B-Instruct}}
& \shortstack[c]{\textbf{Qwen3-Omni} \\ \textbf{30B-A3B-Thinking}}
& \shortstack[c]{\textbf{Stage 1} \\ \textbf{Mixed Bi-modal SFT}}
& \shortstack[c]{\textbf{Stage 2} \\ \textbf{Mixed-Modality RLVR}}
& \shortstack[c]{\textbf{Stage 3} \\ \textbf{Self-Distillation SFT}} \\
\midrule
\textbf{Daily-Omni} & 46.86 & 51.51 & 57.65 & 70.65 & 55.39 & 66.83 & 61.74 \\
\textbf{IntentBench} & 44.06 & 51.06 & 57.36 & 65.38 & 49.65 & 64.04 & 59.99 \\
\textbf{Video-Holmes} & 28.65 & 31.37 & 42.44 & 53.63 & 40.72 & 55.91 & 52.42 \\
\textbf{WorldSense} & 37.17 & 40.28 & 43.83 & 51.27 & 40.42 & 48.85 & 46.79 \\
\textbf{OmniBench} & 37.59 & 43.10 & 48.29 & 54.87 & 41.39 & 56.31 & 54.46 \\
\textbf{UNO-Bench} & 27.25 & 30.46 & 41.11 & 52.17 & 31.36 & 38.78 & 36.87 \\
\textbf{CG-AV-Counting} & 12.73 & 15.13 & 18.57 & 20.28 & 13.83 & 18.62 & 18.09 \\
\textbf{OmniVideoBench} & 35.80 & 33.70 & 38.50 & 39.02 & 32.60 & 37.30 & 37.80 \\
\textbf{AV-Odyssey} & 29.00 & 30.16 & 32.61 & 40.02 & 26.87 & 27.55 & 30.54 \\
\midrule
\textbf{Macro Avg.} & 33.23 & 36.31 & 42.26 & 49.70 & 36.91 & 46.02 & 44.30 \\
\textbf{Original-Weighted Avg.} & 34.65 & 37.76 & 43.05 & 50.99 & 37.81 & 46.12 & 44.94 \\
\bottomrule[1.2pt]
\end{tabular}}
\end{center}

\section{Source Uni-modal Benchmark Pools for the Regression Analysis}
\label{appendix:regression_source_pools}

These tables report the published source scores used to compute the ``Average Vision Performance'' and ``Average Audio Performance'' axes in the Section~3 regression analysis. The model scores are taken from the Qwen2.5-Omni~\cite{xu2025qwen25omnireport} and Qwen3-Omni~\cite{xu2025qwen3omnireport} reports, and the benchmark columns cite the corresponding public benchmark definitions below. They are included only to document the uni-modal reference pools behind the plots and are not additional evaluations of our staged post-training variants.

\subsection{Vision and General Benchmarks}

\begin{center}
\centering
\captionof{table}{Published source scores used to compute the average vision/general-performance axis for the regression analysis in Section~3. Scores are sourced from the Qwen2.5-Omni~\cite{xu2025qwen25omnireport} and Qwen3-Omni~\cite{xu2025qwen3omnireport} reports. Benchmark columns refer to MMMU~\cite{yue2024mmmu}, MMMU-Pro~\cite{yue2024mmmupro}, MathVista~\cite{lu2023mathvista}, MathVision~\cite{wang2024mathvision}, AI2D~\cite{kembhavi2016diagram}, ChartQA~\cite{masry2022chartqa}, MMStar~\cite{chen2024mmstar}, MM-MT-Bench~\cite{agrawal2024pixtral}, and Video-MME~\cite{fu2024videomme}.}
\label{tab:transposed_comparison}
\resizebox{\textwidth}{!}{%
\begin{tabular}{lccccccccc}
\toprule[1.2pt]
Method & MMMU & MMMU-Pro & MathVista & MathVision & AI2D & ChartQA & MMStar & MM-MT & Video-MME \\
\midrule
Qwen2.5-Omni-3B & 53.1 & 29.7 & 59.4 & 20.8 & 79.5 & 82.8 & 55.7 & 5.0 & 62.0 \\
Qwen2.5-Omni-7B & 59.2 & 36.6 & 67.9 & 25.0 & 83.2 & 85.3 & 64.0 & 6.0 & 64.3 \\
Qwen3-Omni-30B-A3B-Instruct & 69.1 & 57.0 & 75.9 & 56.3 & 85.2 & 86.8 & 68.5 & 7.4 & 70.5 \\
Qwen3-Omni-30B-A3B-Thinking & 75.6 & 60.5 & 80.0 & 62.9 & 86.1 & 89.5 & 74.9 & 8.0 & 69.7 \\
\bottomrule[1.2pt]
\end{tabular}%
}
\end{center}

\subsection{Audio}

\begin{center}
    \centering
    \captionof{table}{Published source scores used to compute the average audio-performance axis for the regression analysis in Section~3. Scores are sourced from the Qwen2.5-Omni~\cite{xu2025qwen25omnireport} and Qwen3-Omni~\cite{xu2025qwen3omnireport} reports. Benchmark columns refer to SD-QA~\cite{faisal2021sdqa}, MMSU~\cite{wang2025mmsu}, OpenBookQA~\cite{mihaylov2018openbookqa}, IFEval~\cite{zhou2023ifeval}, AdvBench~\cite{zou2023universal}, VoiceBench~\cite{chen2024voicebench}, and MMAU~\cite{sakshi2024mmau}.}
    \label{tab:omni_benchmark_comparison}
    \resizebox{\textwidth}{!}{
    \begin{tabular}{lccccccc}
        \toprule[1.2pt]
        Method  & SD-QA & MMSU & OpenBookQA & IFEval & AdvBench & VoiceBench Avg & MMAU Avg \\
        \midrule
        Qwen2.5-Omni-3B  & 49.37 & 50.23 & 74.73 & 42.10 & 98.85 & 68.81 & 63.30 \\
        Qwen2.5-Omni-7B  & 55.71 & 61.32 & 81.10 & 52.87 & 99.42 & 74.12 & 65.60 \\
        Qwen3-Omni-30B-A3B-Instruct  & 76.9 & 68.1 & 89.7 & 77.8 & 99.3 & 85.5 & 77.5 \\
        Qwen3-Omni-30B-A3B-Thinking  & 78.1 & 83.0 & 94.3 & 80.6 & 97.2 & 88.8 & 75.4 \\
        \bottomrule[1.2pt]
    \end{tabular}
    }
\end{center}

\end{document}